# Jacobian deformation ellipsoid and Lyapunov stability analysis revisited

Franz Waldner[a*], Rainer Klages[b]

[a] *Physics Institute, University of Zurich, Winterthurerstr. 190, CH-8057 Switzerland*
[b] *Queen Mary University of London, School of Mathematical Sciences, Mile End Road, London E1 4NS, UK*

___________________________________________________________________________________

**A B S T R A C T**
___________________________________________________________________________________

The stability analysis introduced by Lyapunov and extended by Oseledec is an excellent tool to describe the character of nonlinear *n*-dimensional flows by *n* global exponents if these flows are stable in time. However, there are two main shortcomings: (a) The local exponents fail to indicate the origin of instability where trajectories start to diverge. Instead, their time evolution contains a much stronger chaos than the trajectories, which is only eliminated by integrating over a long time. Therefore, shorter time intervals cannot be characterized correctly, which would be essential to analyse changes of chaotic character as in transients. (b) Moreover, although Oseledec uses an *n* dimensional sphere around a point $\underline{x}$ to be transformed into an *n* dimensional ellipse in first order, this local ellipse has yet not been evaluated. The aim of this contribution is to eliminate these two shortcomings. Problem (a) disappears if the Oseledec method is replaced by a frame with a 'constraint' as performed by Rateitschak and Klages (RK) [Phys. Rev. E 65 036209 (2002)]. The reasons why this method is better will be illustrated by comparing different systems. In order to analyze shorter time intervals, integrals between consecutive Poincaré points will be evaluated. The local problems (b) will be solved analytically by introducing the symmetric 'Jacobian deformation ellipsoid' and its orthogonal submatrix, which enable to search in the full phase space for extreme local separation exponents. These are close to the RK exponents but need no time integration of the RK frame. Finally, four sets of local exponents are compared: Oseledec frame, RK frame, Jacobian deformation ellipsoid and its orthogonal submatrix.

___________________________________________________________________________________

[*] Corresponding author: Fax +41 44 635 57 04, Tel. +41 44 923 46 27
*E-mail address*: waldner@physik.uzh.ch; ef.waldner@swissonline.ch; r.klages@qmul.ac.uk

## 1. Introduction

### 1.1 Lyapunov method revisited

Lyapunov exponents [1] are a well-established tool [2-9] to analyse the type of chaos displayed by a trajectory $\underline{x}(t)$ as a solution of a nonlinear *n*-dimensional equation in the phase space $\boldsymbol{R}\{\underline{x}\}$,

$$d\underline{x} / dt = \underline{X}(\underline{x}) \quad (1)$$

Their evaluation is based on proofs by Oseledec [2] that a vector $\underline{u}$ to any arbitrarily chosen neighbouring point of $\underline{x}$ (except the instantaneous direction along $d\underline{x}$ as discussed later) will rotate to the direction of integrated extreme expansion. After an initial period, this local direction obtained through integrated rotation is an inherent feature of each point on the trajectory. Moreover, this local uniqueness applies to the set of all *n* directions of a frame orthogonalised subsequently after each time step. The integration for infinite time $t \rightarrow \infty$ of the corresponding local exponents results in a set of *n* Lyapunov exponents assessing the character of chaos in terms of dynamical instability of the given dynamical system.

The fact that an initial frame rotates after a short time to a locally unique orientation has fascinated many authors. This fascination includes also the corresponding local exponents which has, probably, prevented one to ask: Is the local exponent leading to the largest Lyapunov exponent already locally assessing the strongest expansion, i.e. the strongest divergence of neighbouring trajectories, thus identifying the places where trajectories 'break away', the origin of instability and chaos? Even a quick look at this exponent for the Lorenz model reveals that this is not the case resulting in misleading results for shorter time intervals.

The origin of this discrepancy is inherent to the idea of Lyapunov and Oseledec: They considered a sphere of neighbouring points. Each vector $u$ not orthogonal to the direction d$x$ of the trajectory contains a component parallel to d$x$. This component measures only the expansion along the trajectory, hence the acceleration, which is not connected to instability. Only the component orthogonal to d$x$ analyzes the change of the distance to neighbouring trajectories and can indicate where trajectories start to diverge. This feature is obscured by the admixing of the acceleration. Since in many cases the average acceleration is zero, the integration for long times cancels this contribution, thus only the value for divergence remains.

These shortcomings can easily be avoided by using a method introduced by Rateitschak and Klages (RK) [10]: Select as an initial direction $u$ the direction of d$x$, thus $u$ is parallel to the trajectory. Certainly, this direction will remain parallel to the trajectory for all times. The integral of the corresponding exponent will only describe the average acceleration which for many cases is zero, i.e., if the flow remains bounded and does neither shrink nor blow up. This exponent is not related to the problem of instability. However, all remaining directions of the initial frame, subsequently orthogonalised and their rotation integrated, are describing the divergence of neighbouring trajectories. Their corresponding exponents are the important numbers for analysing the nature of chaos. After an initial period much shorter than for the Oseledec method these directions are unique to the point $x(t)$. The first in the orthogonalizing procedure of the corresponding local exponents is the largest one quantifying where trajectories diverge. Moreover, integrals for shorter time intervals analyse the chaotic character of the trajectories thus discriminating between intervals of strong and weak or no chaos.

This problem of finding coordinate systems in which trivial eigendirections are eliminated has already been addressed by Eckhardt and Wintgen in 1991 [11]. However, they focussed on periodic orbits in conservative two degree of freedom Hamiltonian systems for which they eliminated the two trivial neutral directions along the orbit and perpendicular to it on the energy shell by also requiring certain smoothness properties. Their Hamiltonian method was reviewed and further amended by Gaspard for calculating local Lyapunov exponents and local stretching rates [12]. Other simple numerical methods for computing local Lyapunov exponents and stretching rates have been proposed and tested by Dellago and Hoover [13] and by Rateitschak and Klages (RK) [10].

Both Oseledec and RK methods will be illustrated for the Lorenz [14] and the Rössler [15] model by comparing them with each other and by also applying them to shorter time intervals for transient chaos. It is one of the main points of this contribution to show that the method using a set orthogonal to the flow is more adequate to describe the chaotic behaviour than the Oseledec method using general directions, thus explaining why Rateitschak and Klages [10] found much improved results introducing this method for complex chaotic behaviour.

*1.2 Jacobian deformation ellipsoid*

In order to understand the definition and use of what later on we call 'Jacobian deformation ellipsoid' we first briefly review the origins of Lyapunov instability analysis:

(i) The basic idea of Lyapunov exponents is to follow the evolution of points close to the points $x(t)$ on a trajectory. First, these neighbouring points are chosen on an *n*-dimensional sphere around the starting point $x(t_o)$. Then this sphere is continuously deformed during the time evolution. After a sufficiently long integration time $\tau$ this deformed object is analysed. From the largest expansion direction the largest Lyapunov exponent is evaluated. Starting with the direction of this largest expansion, further orthogonal directions are used to measure their expansions, which complete the set of *n* Lyapunov exponents.

(ii) In principle, the evolution of the deformation should be numerically computed for an infinite number of points on the initial sphere, a tremendous task even for large computers.

(iii) However, the works of Lyapunov [1] and Oseledec [2] propose a well-established, much less elaborate method: Arbitrarily defined at the start, choose an orthogonal frame of only *n* directions, follow their evolution - orthogonalised always in the same order by the Gram-Schmidt orthonormalisation method before each integration step - and measure their expansion, thus disregarding their rotation. The Lyapunov exponents are then obtained as the averages of the logarithms of these expansions.

It would be interesting to test to which extent (iii) corresponds to (i) in case (ii) could be treated with less numerical effort. It is the aim of this section to propose a simpler method.

Before using the computer for doing (ii), let's go back to the 19th century of Jacobi. The main idea was then: Think of a reduction of the problem and retain only the essentials. Jacobi [16] used only the first derivative to define his matrix $J$ at a point $\underline{x}$,

$$J(\underline{x}) = d\underline{X} / d\underline{x} \qquad (2)$$

.
He realised that the sum of the diagonal elements of $J$ measures the local rate of change of the volume of a sphere around $\underline{x}$ in the limit of infinitesimally small radius. However, nothing has been said then about the deformation of this sphere. Although finally global quantities should be evaluated, we will first consider 'instantaneous local' quantities defined by the deformation between a fixed time $t$ and $t+dt$, starting as a sphere at $t$.

Applying the 19th century method to the deformed sphere, what is the first approximation? This has been shown already by Farmer et al. [4] with a picture of an ellipse in two dimensions. Green and Kim [7] describe a general ellipsoid for $n$ dimensions, which is continuously deformed in time but remains always an ellipsoid. The problem (i) is then solved by using the $n$ principal axes and their corresponding expansions for the final ellipsoid. Obviously, these principal axes are orthogonal in accord with (iii), since the ellipsoid has inversion symmetry. An $n$-dimensional ellipsoid is described by a symmetric n x n matrix $E$, with orthogonal eigenvectors as principal axes and real eigenvalues, and is determined by $n(n+1)/2$ parameters. At this point it seems worthwhile to note that $E$ yields the radius for the infinite number of possible directions as diagonal elements $(E_U)_{ii}$ of $E_U$, after transforming to a new coordinate system with the unitary matrix $U$ and its transpose $U^T$ as $E_U=U^T E U$. The matrix $E$ has the same size as $J$. However, $J$ is in principle not symmetric causing complex eigenvalues and non-orthogonal complex eigenvectors. Furthermore, its diagonal elements are connected with the rate of change of the radius of a sphere, not with the radius of the ellipsoid described by $E$. Therefore, why not try to symmetrise $J$ while keeping the diagonal elements by defining a symmetric matrix $S = ½ (J + J^T)$ with orthogonal eigenvectors and real eigenvalues in order to describe the rates of change of the radii when the sphere is transformed in first order into an ellipsoid? This paper aims to convince the reader that $S$ is exactly describing in first order the 'instantaneous local rates of exponential stretching ratio' – in short 'local exponents' - for all possible directions, although only $n^2$ numbers are involved in $J$.

This symmetric result has almost been found by Greene and Kim [7] in their eqs. (26)-(28). They showed that their instantaneous local expansion exponents $\lambda_U$ along any orthogonal set of directions $U$ are given by the diagonal elements $K_{ii}$ of $K=U^T J U$ (note that in general $J$ and $K$ are not symmetric). Simply reduce this equation for only one direction $\underline{u}$ to the scalar product $\lambda_{\underline{u}}=(\underline{u},J\underline{u})$. This form can be derived in a direct geometrical way providing probably one of the simplest approaches to define local Lyapunov exponents. The derivation is so short that it will be sketched here in the introduction, as follows:

The local exponent $\lambda$ for a dynamical system given by eq.(1) can be defined as $\lambda= (1/\Delta t) \ln (r)$, with $r = (d+\Delta d)/d = 1 + \Delta d/d$ being the stretching ratio for the length $d$ of a vector $\underline{u}$ pointing from a point $\underline{x}(t)$ towards a neighbouring point. Using the Jacobian matrix $J$, the new vector after time $\Delta t$ is found in first order to be $\underline{u}(t+\Delta t) = \underline{u}(t)+J \underline{u}(t) \Delta t$. This new vector has rotated and changed its length to $d+\Delta d$. It is now important to eliminate the rotation by projecting the change $\Delta \underline{u} = J \underline{u} \Delta t$ onto $\underline{u}$ by using the scalar product $\Delta d = (\underline{u},J\underline{u}) \Delta t / |\underline{u}|^2$. valid for $\Delta t \to 0$. Putting this into $r = 1 + \Delta d/d$ and expanding $\ln(1+\alpha) \approx \alpha$ for $|\alpha|<<1$ results in $\lambda= (1/\Delta t) \ln (r) \approx (1/\Delta t) (\underline{u},J\underline{u}) \Delta t / |\underline{u}|^2$. The simple result $\lambda=(\underline{u},J\underline{u})$ is then found if the vector $\underline{u}$ has been normalised to unity. Note that the `logarithmisation' is no longer visible in the form for $\lambda$ after the above approximation has been applied. Furthermore, note that this result has the correct dimension of reciprocal time as needed for a Lyapunov exponent defined above as 'the rate of exponential stretching ratio'. Interestingly, the form $\lambda_{\underline{u}}=(\underline{u},J\underline{u})$ shows that $\underline{u}$ and $-\underline{u}$ give the same value implying inversion symmetry. In more detail, the elements $J_{ik}$ of the matrix $J$ appear in pairs $(J_{ik} + J_{ki})$ in the scalar product $(\underline{u},J\underline{u})$. Hence $J$ can be replaced by the symmetric matrix $S = ½ (J + J^T)$, with $J^T$ denoting the transposed matrix. This gives the same value for $\lambda_{\underline{u}}= (\underline{u},S\underline{u})$ as $\lambda_{\underline{u}} = (\underline{u},J\underline{u})=(\underline{u},J^T\underline{u})$.

Going now back to the above form $K=U^T J U$ of Green and Kim [7] and replacing there $J$ by $S$, the diagonal elements $T_{ii}$ of the new matrix $T=U^T S U$ are equal to $K_{ii}$. In contrast, this new $T$ is a symmetric matrix corresponding to the symmetrised $K$ with $T = ½ (K + K^T)$. Furthermore, the transformation of $J$ with $K=U^T J U$ could be interpreted as transforming $J$ into the new reference system $U$ and the result might be denoted by $J_U$. Similarly, $T =U^T S U$ transforms $S$ into $S_U= U^T S U$ with its diagonal elements $(S_U)_{ii}$ as instantaneous local exponents in the $U$ directions. Thus $S$ is the generating form containing all deformations of the ellipsoid and will be called 'Jacobian deformation ellipsoid' (not to be confused with 'deformed ellipsoid *E*' which describes directly the radii, not the rates of their changes). Note that using the symmetry argument is a consequence of the reduction of the deformation to the lowest order. In reality, the deformations are more complicated than being captured in lowest order. As a by-product, the largest eigenvalue of the main principal axis of *S* gives the extreme rate of expansion of the sphere, its eigenvectors rarely being parallel to any of the frames in Section 1.1

The above first step defines the instantaneous local deformation. Before doing the second step, let us recall here three relevant features of the 'Jacobian deformation ellipsoid' *S*: First, the matrix $S = \frac{1}{2}(J + J^T)$ is obviously *symmetric* by definition. Second, the `logarithmisation' is not visible explicitly since the approximation $\ln(1+\alpha) \approx \alpha$ for small $\alpha$ has been used; nevertheless, *S* describes the *exponential rates* of instantaneous local stretching. Third, *S* yields this rate for *all directions* around the point $\underline{x}$, although only $n^2$ numbers of the elements of the local Jacobian matrix $J(\underline{x})$ are needed to define the matrix *S*.

The second step now consists of evaluating global quantities both for schemes (i) and (iii) above in order to test them in the long-time limit. In both cases, global exponents are defined as final deformations after integrating all deformations during a time interval $\tau = t_{final} - t_{initial}$ for $\tau \to \infty$.

Before going into further detail, two important questions have to be answered: Have `deformations of previous deformations' to be evaluated? And have rotations of the ellipsoids to be incorporated?

The answers to both questions are illustrated in figure 1 of Benettin et al. [3]: After each time step the new deformation does not account for the previous deformation, that is, the deformation of the previous deformation is of higher order and can be neglected. Thus instantaneous local quantities can be averaged if integration is replaced by small but finite time steps as in numerical work. As far as rotation is concerned, each new start in Benettin's figure 1 is from a rotated direction. Since *S* does not contain any information about the rotation, it has to be incorporated separately, thus at each time step the transformed $S_U(t) = U^T(t) S(t) U(t)$ has to be evaluated. Note that for $U(t)$ any of the two orthonormalised frames of section 1.1 could be used. Furthermore, note that $S_U(t)$ has usually non-zero off-diagonal elements implying that the diagonal elements are no eigenvalues.

Now, on the basis of the new Jacobian deformation ellipsoid *S*, differences between (i) and (iii) will be described. The evaluation according to (i) has to be done in two steps: First, average all local matrices $S_U(t)$ along a typical trajectory. Obviously, each of the $n^2$ elements of $S_U(t)$ has to be averaged separately. Hence this averaging results in a final symmetric matrix $S_{final}$. In a second step, only for this final matrix $S_{final}$ the eigenvalues and the eigenvectors have to be evaluated corresponding to global Lyapunov exponents and Lyapunov directions, respectively.

The recipe for performing (iii) is much simpler: The averages of the *n* diagonal elements $(S_U)_{ii}(t)$ of the local matrix $S_U(t)$ along a typical trajectory are the global Lyapunov exponents. Again obviously, each of the *n* diagonal elements of $(S_U)_{ii}(t)$ has to be averaged separately. Comparing both methods, the results of (iii) are already incorporated in the final matrix $S_{final}$ as diagonal elements. Hence, to test the equivalence between (i) and (iii) it is sufficient to compare the values of the diagonal elements of $S_{final}$ with its eigenvalues. This test has been performed numerically for the 'stable' Lorenz chaos. There are small but distinct deviations for the free running rotation of Oseledec. The test is successful for the constraint rotation of RK.

So far the literature has mainly focused on local and global Lyapunov exponents. Both quantities have been defined above within a new approach, and tests of this concept will be described later on in this paper. So what else? Previous concepts of Lyapunov instability are only appropriate for 'stable' chaos, where a trajectory does not change its character in time. They are not suitable to analyse transients, crises, or continuous changes of parameters in time in equations of motion. However, an adequate method is easy to find: Instead of only considering 'global quantities' defined for an infinite interval of time $\tau$, try a series of successive finite time intervals $\tau_n$, each starting at $t_n$ and ending at $t_{n+1}$. It is essential to define the successive times $t_n$ such that the resulting data correspond to the character of the trajectory. To give an example, for a 'stable' Lorenz chaos time intervals $\tau_n$ between successive Poincaré points are convenient, which have been chosen such that each interval describes a loop on the left or on the right hand side of the strange attractor. Again both methods (i) and (iii) are compared for different frames of rotation. Only the RK frame produces rather small but distinct deviations of the values for (i) and (iii). These discrepancies are caused by residual acceleration parts parallel to the flow $d\underline{x}$. It is easy to eliminate all these directions by forming the $n-1$ dimensional subspace $S_\perp^{(2...n)}$ orthogonal to the flow of each instantaneous local $S_U(t)$ with *U* as the RK frame, a straightforward evaluation. For (i) the eigenvalues of the averaged subspace are then compared with (iii) as the averages of the exponents orthogonal to $d\underline{x}$ of RK in the same time interval. Indeed, both values are equal within numerical accuracy for all time intervals.

The method (iii) based on a series of finite intervals is then successfully applied to a transient Lorenz chaos. It distinguishes well between an initial period of weak chaos, an intermediate period of strong chaos, and a final spiralling onto a stable fixed point. Also here the RK frame is superior to the Oseledec frame.

As a by-product, each instantaneous local subspace $S_\perp^{(2...n)}$ - to be calculated at any point $x$ of the phase space $R\{x\}$ – has $n$-1 eigenvalues. The largest eigenvalue within this subspace accounts for the extreme rate of divergence between neighbouring trajectories, not obscured by partial acceleration. This yields a novel indicator for extreme local divergence, which can be used to find 'hot spots' of maximum local dynamical instability in the whole phase space.

These applications demonstrate that the new 'Jacobian deformation ellipsoid' is a powerful tool, which furthermore opens a pedestrian approach to defining both local and global Lyapunov exponents.

*1.3 Comparison of four types of instantaneous local exponents*

The novelty of this article is that it compares four types of local directions and exponents with each other. The first two are strictly local to $x$. They use only the knowledge of the local Jacobian $J$ and the value of d$x$, there is no need to evaluate any trajectory by integration:

1. The instantaneous local extreme expansion exponent of a local sphere and its corresponding direction are found as the largest eigenvalue and eigenvector of $S$. Note that this yields the maximal possible value of the exponent for all four different methods, and the direction of this maximal deformation is rarely parallel to the direction obtained by the other three methods.

2. The extreme exponent for the instantaneous local divergence of neighbouring trajectories measured orthogonal to d$x$ and its direction are found as the largest eigenvalue and eigenvector of the subset $S_\perp$. Note that this exponent is the maximal possible value of the largest exponent for divergence in what we will call the $W$ frame of RK as described in method 4 below.

The other two types are evaluated by method (iii). and need integrations both of a trajectory and the directions of the frames.

3. The standard method of Oseledec (O) with an integrated free running frame called $V$ frame produces $n$ instantaneous local exponents and directions.

4. The method of RK using an integrated $W$ frame, where the first direction is always constrained to be parallel to d$x$ thus assessing the instantaneous local acceleration parallel to d$x$ (needs no integration). The remaining instantaneous local exponents are orthogonal to d$x$ describing the divergence of neighbouring trajectories, with their corresponding directions found by integration of the $W$ frame.

*1.4 A test of how well the exponents of methods 3. and 4. are correlated with the exponents of methods 1. and 2.*

A heuristic test uses Poincaré points and the corresponding local exponents. The distribution of these local exponents is tested as a function of the distance between the corresponding Poincaré points. The Lorenz model reveals that the O method has a much larger chaotic character than the RK method. This confirms again that the RK method is more efficient than the O method. Only the RK method is adequate to evaluate meaningful Lyapunov exponents, which furthermore indicate where the instabilities start locally.

## 2. Treating a local point $x$ of the phase space

*2.1 Geometric interpretation of local expansion*

Local expansion of a vector $u$ to a neighbouring point of $x(t)$ during a time interval $\Delta t$ will be described geometrically. In first order, using the Jacobian matrix $J$

$$u(t+\Delta t) = u(t) + J(t)\, u(t)\, \Delta t\ =\ u(t) + \Delta u \quad (3)$$

the new vector $u(t+\Delta t)$ will have rotated and changed its length from $d$ to $d+\Delta d$. If only the change $\Delta d$ of the length is considered, the rotation can be eliminated by projecting $\Delta u$ onto $u$ by using the scalar product $\Delta d \approx (u, Ju)\, \Delta t / |u|^2$ valid for $\Delta t \to 0$. With the ratio $r$ of stretching $r = (d+\Delta d) / d = 1 + \Delta d/d$ the local exponent $\lambda_u$ for the direction of $u$ is found according to the form, see Greene and Kim [7],

$$\lambda_u = (1/\Delta t)\, \ln(r) \qquad (4)$$

giving $\lambda_u = (1/\Delta t) \ln(1 + \Delta d/d) = (1/\Delta t) \ln[1 + (u, Ju)\, \Delta t / |u|^2]$.

Using $\ln(1+\alpha) \approx \alpha$ for $|\alpha| \ll 1$, the result is the scalar product

$$\lambda_u = (u, Ju) \qquad , \qquad (5)$$

if the vector $u$ has been normalized to unity.

Greene and Kim [7] published in their eqs. (26-28) the same result in form of the diagonal elements $K_{ii}$

$$\lambda_i = K_{ii} \qquad K = V^T J V \qquad (6)$$

for an orthonormalised reference frame $V$ and the Jacobian matrix $J$.

Note that in the scalar product $(\underline{u}, J\underline{u})$ the values $J_{ik}$ and $J_{ki}$ appear in pair sums $(J_{ik} + J_{ki})$. This symmetry is enforced by the special structure of the scalar product $(\underline{u}, J\underline{u})$: on both sides there occurs the same vector $\underline{u}$. Therefore, the same result is obtained for a symmetrised Jacobian matrix $S$ constructed by adding the transpose $J^T$.

$$S = \tfrac{1}{2}(J + J^T) \qquad (7)$$

$$\lambda_{\underline{u}} = (\underline{u}, S\underline{u}) \qquad (8)$$

In more detail, the Jacobi matrix $J$ can be decomposed into $J = S + D$ with the symmetric matrix $S$ defined by eq. **(7)** and the anti-symmetric matrix $D$ with $D_{ik} = -D_{ki}$ defined by the difference $D = \tfrac{1}{2}(J - J^T)$. The scalar product of eq. (5) thus results in $(\underline{u}, J\underline{u}) = (\underline{u}, S\underline{u}) + (\underline{u}, D\underline{u})$. The second term $(\underline{u}, D\underline{u})$ is zero, because the same vector on both sides of the scalar product enforces terms which add up to zero due to the anti-symmetry of $D$. Therefore, the reduction of $(\underline{u}, J\underline{u})$ to $(\underline{u}, S\underline{u})$ as performed in eq. (8) is justified. Furthermore, note that $(\underline{u}, D\underline{u}) = 0$ implies that the vector $D\underline{u}$ is orthogonal to $\underline{u}$ describing only a rotation of $\underline{u}$. At first glance, for capturing the rotation of $\underline{u}$ in eq. (3) it is thus tempting to reduce the term $J\underline{u}$ therein to $D\underline{u}$. However, the vector $S\underline{u}$ can have any direction thus incorporating both elongation and rotation before it is projected onto $\underline{u}$ by the scalar product $(\underline{u}, S\underline{u})$ to find only elongation. Hence, for describing the full rotation of $\underline{u}$ it is essential to keep both terms $(S+D)\underline{u}$ in eq. (3).

The symmetrised Jacobian matrix $S$ will pave the way to identify the Jacobian deformation ellipsoid as demonstrated in the next section.

## 2.2 The Jacobian deformation ellipsoid

A local $n$ dimensional sphere around a point $\underline{x}$ of eq. (1) will be rotated and deformed after a time interval $\Delta t$ into a complicated geometrical object. Only in linear approximation its form can be described by a symmetric $n$ dimensional ellipsoid with orthogonal principal axes of the principal deformation exponents. Here, a straightforward simple way to find this ellipsoid will be described: The expansion exponent $\lambda_{\underline{u}}$ in any direction $\underline{u}$ is found by the scalar product $(\underline{u}, S\underline{u})$, see eq. (8). Defining an orthonormal set $U$ and its transpose $U^T$, the transformed matrix $S_U = U^T S U$ can be evaluated. It is now important to note that the scalar product of eq. (8) implies that only the diagonal elements $(S_U)_{ii}$ are equal to the local expansion exponents $\lambda_i$ corresponding to the $n$ orthogonal normalised directions $\underline{u}_i$ in $U$. Note further that the off-diagonal elements $(S_U)_{ik}$ with $i \neq k$ can have any non-zero value.

Let us first consider the special case that all off-diagonal elements $(S_U)_{ik}$ with $i \neq k$ are zero. In this case, the diagonal elements $(S_U)_{ii} = \lambda_i$ could be eigenvalues of $S$. Since $S$ is by definition symmetric, its eigenvalues $\alpha_i$ are real and the corresponding eigenvectors $\underline{a}_i$ are both real and orthogonal. It is then obvious from eq. (8) that the eigenvalues $\alpha_i$ can indeed be local expansion exponents. By expanding the arbitrary vector $\underline{u}$ in eq. (8) into the basis of eigenvectors of $S$, in complete analogy to the expectation value problem of a quantum mechanical operator [17], one concludes that the eigenvalues $\alpha_i$ are the extreme local expansion exponents, denoted by $\lambda_i^{(e)}$. This implies that the eigenvectors $\underline{a}_i$ correspond to the extreme vectors $\underline{u}_i^{(e)}$ of $U^{(e)}$, which are orthogonal by definition. They transform the matrix $S$ into its diagonal eigenvalue form $S_{\mathrm{diag}} = (U^{(e)})^T S U^{(e)}$. The vectors $\underline{u}_i^{(e)}$ thus define the principal axes of the ellipsoid into which the local sphere is deformed.

Note that in the general case of arbitrary directions of $U$ the corresponding local expansions exponents $\lambda_i = (S_U)_{ii}$ are not eigenvalues of $S$. In summary, according to eq. (8) the matrix $S$ can be considered as the generator of deformations in all possible local directions including the principal axes which yield the extreme values [18]. It is therefore justified to call $S$ the $n$-dimensional 'Jacobian deformation ellipsoid'. Note that $S$ does not describe the deformed sphere. It accounts for the rate of expansion (positive values) or contraction (negative values) in any direction; hence, this 'deformation ellipsoid' can have values of both signs.

It will be convenient to order the exponents $\alpha$ according to their values $\alpha_k > \alpha_{k+1}$, thus the largest first, and the eigenvectors $\underline{a}_k$, written as columns in the matrix $A_x$ accordingly, its components expressed in the frame of $R\{\underline{x}\}$.

The Jacobian deformation ellipsoid can be written as a symmetric $n$ dimensional tensor $T$, its explicit form depending on the frame of reference. The simplest way is in the local frame of principal axes, the matrix $T_{diag}$ with only the exponents $\alpha_k$ in the diagonal. Its trace as the sum of the exponents is clearly the rate of change of the volume of the sphere around $\underline{x}$. The more convenient form $T_{\underline{x}}$ would be expressed in the reference frame of $R\{\underline{x}\}$, to be found by back transformation, resulting, obviously, in the form of $S$ expressed usually in the reference frame of $R\{\underline{x}\}$, with the trace unchanged as the rate of change of the volume,

$$T_{\underline{x}} = A_{\underline{x}} T_{diag} A_{\underline{x}}^T = S_{\underline{x}} \qquad (9)$$

An alternative frame of reference will be described in the next section.

*2.3 The orthogonal Jacobian reduced ellipsoid $S_{\perp}^{(2,...n)}$ for divergence*

Although the Jacobian deformation ellipsoid contains all local information about neighbouring points, it is essential to find the principal exponents and their directions orthogonal to the flow. Only these directions indicate true divergence, because they are not disturbed by partial acceleration.

First the reason will be explained why these directions are important. Then an arbitrary local frame of reference is introduced serving to find the local frame of extreme divergence. The ellipsoid is transformed into this frame. After its truncation to the *n*-1 dimensional subspace orthogonal to the flow d$\underline{x}$, the new extreme exponents and principal directions are found by solving the *n*-1 dimensional eigenvalue problem. Although these procedures are straightforward, the matrix operations will be described in more detail in order to facilitate their implementation in programs for Lyapunov exponents.

*2.3.1 The mixing of divergence and acceleration*

The Jacobian deformation tensor will be described now by $S_{\underline{x}}(\underline{x})$ at the point $\underline{x}$, the subscript denotes that the matrix is written in the coordinate system $R\{\underline{x}\}$. $S_{\underline{x}}(\underline{x})$ describes the rate of stretching along all possible directions $\underline{u}$ to a neighbouring point of any of the points $\underline{x}$ of the phase space without the need to execute any integration of eq. (1). Whereas most of these points belong to a neighbouring trajectory, there is one point exactly on the trajectory through $\underline{x}$. This point is found for $\Delta t \rightarrow 0$ along the vector $\underline{X}$ of eq. (1) which describes the flow through $\underline{x}$. Defining a flow unit vector $\underline{f}_\parallel = \underline{X} / |\underline{X}|$, the corresponding expanding exponent $\varphi_\parallel$

$$\varphi_\parallel = (\underline{f}_\parallel, S_{\underline{x}} \underline{f}_\parallel) \qquad (10)$$

does not describe any divergence of a neighbouring trajectory. Instead, it is the relative acceleration (d$v$/d$t$)/$v$ related only to the change of velocity $v$ along the trajectory through $\underline{x}$ [7].

Hence, any vector $\underline{u}$ can be decomposed into a vector sum $\underline{u} = c_\parallel \underline{f}_\parallel + c_\perp \underline{g}_\perp$, where $\underline{g}_\perp$ is the appropriate unit vector in the space orthogonal to $\underline{f}_\parallel$. Therefore, $\underline{u}$ has its expansion exponent $\lambda_{\underline{u}}$ composed of the exponent $\varphi_\parallel$ related only to acceleration and $\lambda_{\perp g}$ describing only the divergence of the neighbouring trajectory line in the $\underline{g}_\perp$ direction,

$$\lambda_{\underline{u}} = c_\parallel^2 (\underline{f}_\parallel, S\underline{f}_\parallel) + c_\perp^2 (\underline{g}_\perp, S\underline{g}_\perp) = c_\parallel^2 \varphi_\parallel + c_\perp^2 \lambda_{\perp g} \qquad (11)$$

*2.3.2 Constructing the orthogonal Jacobian reduced ellipsoid for divergence*

Hence, if the sole interest is to find locally the largest divergence of neighbouring trajectory lines, and not to be disturbed by interference with acceleration, the *n*-1 dimensional subspace orthogonal to the flow direction $\underline{f}_\parallel$ has to be constructed. This will be performed using an arbitrary local orthonormal set of reference $F$ with the first column as the unit vector $\underline{f}_\parallel$. Then, construct the remaining *n*-1 vectors $\underline{f}_k$ by permutation of the components of $\underline{f}_\parallel$. Finally, use a Gram-Schmidt procedure to make the $\underline{f}_k$ orthogonal as columns of the matrix $F_{\underline{x}}(\underline{x})$ expressed in the frame of reference $R\{\underline{x}\}$ (denoted by subscript $\underline{x}$) at the local point $\underline{x}$.

The Jacobian deformation ellipsoid $S_{\underline{x}}$ is transformed into the $F$ frame by

$$S_F = F_{\underline{x}}^T S_{\underline{x}} F_{\underline{x}} \qquad (12)$$

The first row and column of $S_F$ describe the expansion along the flow. The remaining *n*-1 dimensional reduced square submatix $S_\perp^{(2,...,n)}$ contains all expansions in the orthogonal subspace.

*2.3.3 The principal local exponents for divergence*

The reduced square submatix $S_\perp^{(2,...n)}$ is symmetric and has *n*-1 eigenvalues $\beta_{\perp k}$ and eigenvectors $\underline{b}_{\perp k}$ as the principal perpendicular local exponents and directions, respectively, again both to be ordered according to their values $\beta_{\perp k} > \beta_{\perp k+1}$, the largest first, and $B$ is the *n*-1 dimensional matrix containing the ordered vectors $\underline{b}_\perp$ as columns.

It is worthwhile to construct a local reference frame $\{\underline{h}\}$ as the matrix $H$ with the first vector as flow direction $\underline{f}_\parallel$ and the remaining directions as local perpendicular extreme expansion directions $\underline{b}_{\perp k}$. In the $F$ frame, the first row and column are zero except $H_{F11}=1$. The remaining submatrix is filled with the matrix $B$. This frame $H_F$ can be transformed into the reference frame of $R\{\underline{x}\}$ by the arbitrary matrix $F_{\underline{x}}$

$$H_{\underline{x}} = F_{\underline{x}} H_F \qquad (13)$$

Although the exponents $\beta_{\perp k}$ were evaluated already by solving the eigenvalue problem of the reduced submatrix $S_\perp^{(2,...n)}$, a general relation as a numerical control could be written using the frame matrix $H_x$ in a matrix product with the diagonal elements $(...)_{ii}$ giving $\varphi_\parallel$ as the first, and $\beta_{\perp k=1,...,n-1}$ as the remaining numbers

$$\varphi_\parallel = (H_x^T \, S_x \, H_x)_{11}$$
$$\beta_{\perp k = (1,...,n-1)} = (H_x^T \, S_x \, H_x)_{(k+1),(k+1)} \qquad (14)$$

### 2.3.4 Exploring the whole phase space for extreme local divergence

Before a specific trajectory is evaluated, it seems worthwhile to explore the phase space by producing an *n* dimensional map of the principal exponent of local divergence $\beta_{\perp k=1}$ in order to find 'hot' regions of large divergence or 'cool' regions of missing divergence. However, applying this procedure to the Lorenz attractor the interpretation is not trivial; there are strong 'hot' and very 'cool' regions well outside the strange attractor. Examples will be given later.

Moreover, changing the parameters in eq. (1) could result in a very different behaviour, more, stronger, less or no 'hot spots' in the whole phase space. To test this would be very elaborate by evaluating various trajectories. The procedure described here is much faster.

## 3. Exponents following a trajectory

A specific trajectory $x(t)$ is found after choosing a starting point $x_0 = x(t=0)$ by integrating eq. (1) starting at $x_0$. In the spirit of Lyapunov, *n* local exponents could be evaluated if a specific orthogonal set of directions is defined for each $x(t)$. The average of these local exponents will for $t \to \infty$ lead to the Lyapunov exponents $\Lambda_k$ characterizing the type of trajectory.

Therefore, the problem of finding these local directions is essential. First, the well established method of Oseledec (O) without a constraint will be shortly described. Then the new method of Rateitschak and Klages (RK) with their constraint will be introduced. A comparison and a test will how later on that only the second method should be used, implying a fundamental change for the description of instability.

### 3.1 The Oseledec (O) method for the local frame V

According to Oseledec [2], any arbitrarily chosen orthogonal frame $V_0$ at the origin $x_0$, rotated according to eq. (1), then orthogonalised and normalised after each time step $\Delta t$, will become a unique local frame $V[x(t)]$ for each point $x(t)$ of the trajectory after an initial transient time $\tau_{transV}$. The rotation can be evaluated by applying eq. (1) to neighbouring points or by using eq. (3) as proposed by Greene and Kim [7]. The idea is that the first direction, which is never adjusted by the orthogonalising process, will turn to the direction of strongest divergence from the trajectory, independently from its starting direction in $V_0$. The instantaneous local expansion exponents $\lambda_{Vk}$ can be found by eq. (5) or (8), and their averages will be the global Lyapunov exponents $\Lambda_{Vk}$ for $t \to \infty$. The first exponent $\Lambda_{Vk=1}$ will be the largest.

At this point it is interesting to note that and Meier [19] found in their numerical analysis of the angles of the *V* frame that the direction corresponding to the smallest ('most negative') local exponent is always nearly orthogonal to the flow. Small deviations are probably due to finite step integration and numerical limitations.

### 3.2 The Rateitschak and Klages (RK) constraint frame W

Rateitschak and Klages (RK) [10] introduced a new concept for a local frame **W**. The rotation can be made by eq. (3), and the first vector $w_{k=1} = f_\parallel$ is always set parallel to $\dot{X}/|\dot{X}|$. The remaining directions are then orthogonalised always in the same order. Also here, after a transient time $\tau_{transW}$ a unique frame $W[x(t)]$ for each point of the trajectory $x(t)$ will be defined. The local expansion exponents $\lambda_{Wk}$ can be found by eq. (5) or (8) by using **W** instead of **V**, and their time averages will be the global exponents $\Lambda_{Wk}$ for $t \to \infty$. Now, the first $\Lambda_{Wk=1} = \Lambda_\parallel$ will not be the largest. It has a different function: it measures the mean acceleration of the flow, which is zero for many chaotic models. The remaining $\Lambda_{Wk>1} = \Lambda_{\perp Wk}$ all describe only divergence of neighbouring trajectory lines, hence the second $\Lambda_{Wk=2}$ will be the largest.

A numerical test with the Lorenz model [14] confirmed that the third direction of the *V* frame is not only always nearly orthogonal to the flow [20], but also always nearly parallel to the third direction of the *W* frame, with deviations of the order of the deviations within the *V* frame.

### 3.3 Comparing recovery times of Oseledec with Rateitschak and Klages

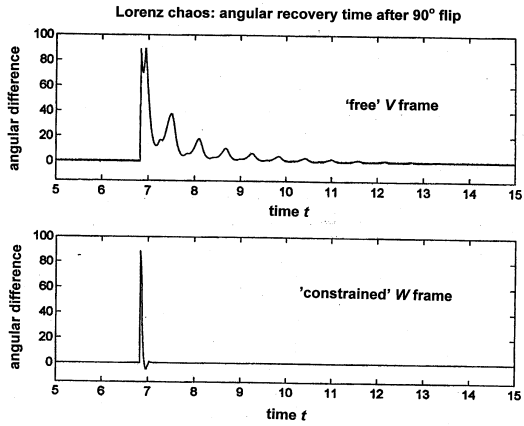

**Fig. 1.** Lorenz chaos: Difference of the angles between the maximal expansion direction in a local frame flipped at time $t=6.8$ and the respective direction in the unflipped frame vs. time $t$. <u>Above</u>: free frame $V$ of O [3]; <u>Below</u>: constraint frame $W$ of RK [10] exhibiting a much shorter recovery time.

How fast does an arbitrarily chosen frame approach the locally unique frame directions, described by a transient time $\tau_{transient}$? This question is analysed in the Lorenz system by measuring the recovery time $\tau_{recover}$ of a suitably rotated frame. The tests starts with a frame $V$ at time $t_0$. At a time $t_1 \gg \tau_{transient}$ a new frame $V_{rot}(t_1)$ is constructed, which is rotated by 90 degrees with respect to $V(t_1)$. Both frames are integrated in time, and the difference of the angles between the maximal expansion direction of the flipped $V_{rot}(t)$ and the unflipped $V(t)$ are plotted in Fig. 1 as a function of time $t$. Clearly, the recovery time $\tau_{recover}$ is much longer for O (top) than for RK (bottom). For RK, the direction of the flow is obviously not rotated at $t_1$. Therefore, only $n$-1 directions have to readjust. These recovery times $\tau_{recover}$ are an indication for the transient times $\tau_{trans}$ after $t_0$.

In addition, it seems worthwhile to note that the exponent of the acceleration of the flow has zero recovery time, since its direction $\underline{X}/|\underline{X}|$ is the local value of eq. (1) for each point $\underline{x}$ of the phase space.

### 3.4 Oseledec vs. Rateitschak and Klages Lyapunov exponents for $t \to \infty$

The corresponding values of the Lyapunov exponents $\Lambda$ for $t \to \infty$ are numerically tested for three and four dimensions using the Lorenz and the Rössler model, respectively. Fig. 2 (Lorenz), and Fig. 3 (Rössler) display the results as functions of integration time $t$. At the bottom, the Lorenz $x$ component or the Rössler $x_3$ component is shown. On top, the free $V$ frame of O is used, below the constraint $W$ frame of RK. Fig. 4 displays in more detail the values between $t = 700$ and $800$ of the first three integrated exponents.

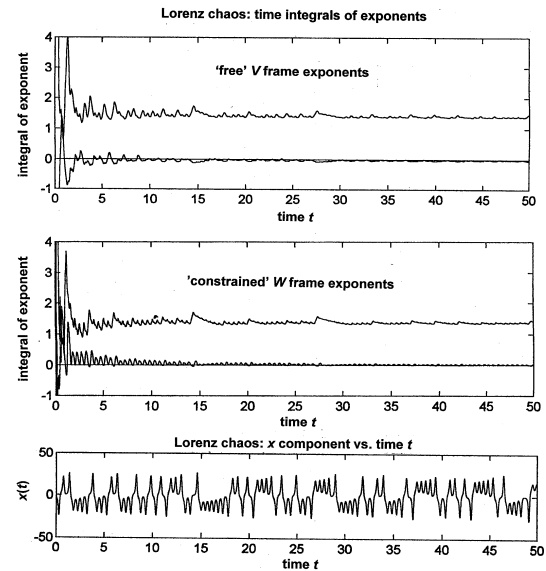

**Fig. 2.** Lorenz chaos: <u>Bottom</u>: $x$-component vs. time. <u>Top</u>: The first two exponents integrated vs. time following the free $V$ frame according to O. [3]. <u>Centre</u>: the same integrated exponents vs. time following the constraint $W$ frame according to RK [10].

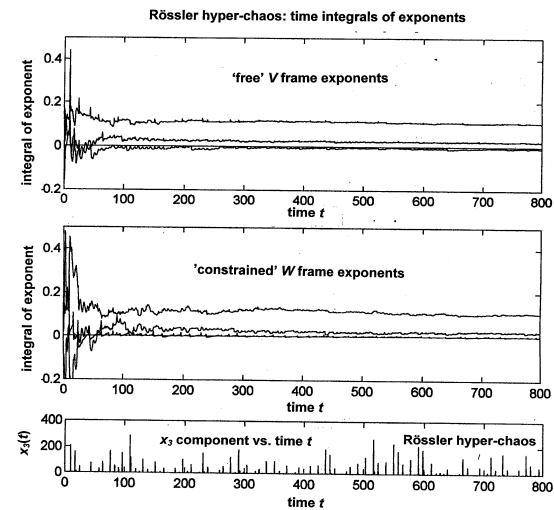

**Fig. 3.** Rössler hyper-chaos: <u>Bottom</u>: $x_3$-component vs. time. <u>Top</u>: The first three exponents integrated vs. time following the free $V$ frame according to O. [3]. <u>Centre</u>: the same integrated exponents vs. time following the constraint $W$ frame according to RK [10].

The first two exponents are positive indicating hyper-chaotic behaviour; the third is zero for zero mean acceleration. The final magnitudes are nearly equal and within the fluctuations of both frames, see Fig.4.

The fluctuations are more pronounced for the $W$ frame (below), since its recovery time is much shorter, therefore the strong peaks of $x_3$ (see Fig. 3 bottom) are not so well integrated out as for the $V$ frame (above). Note that the third exponent of the free $V$ frame (above) is slightly below zero although it should be zero according to the theory.

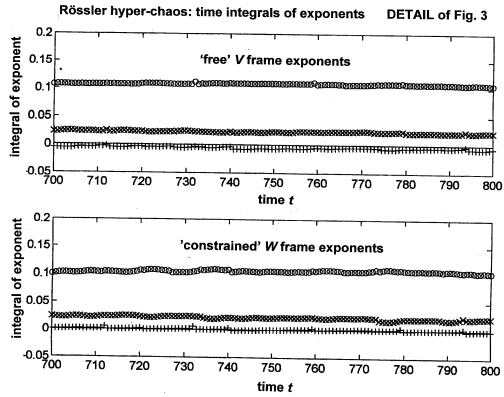

**Fig. 4.** Rössler hyper-chaos: Blow-up of the final time sequence between *t*=700 and 800 of Fig. 3.

These interesting results confirm that the largest Oseledec exponent $\Lambda_{Vk=1}$ describes the divergence, although its local values are mixed with acceleration. For the Lorenz model the second $\Lambda_{Vk=2}$ and for the Rössler model the third exponent $\Lambda_{Vk=3}$ tend to a zero value, thus describing the mean acceleration $\Lambda_{Wk=1}$ of the RK method by averaging out local acceleration and divergence.

Hence, for infinite time and 'stable' chaos both methods are equivalent since the contribution of acceleration is cancelled by the integration. However, as shown later, for transients and chaotic regimes changing their character only the RK method is adequate.

## 4. Comparing local directions and exponents as functions of time *t*

*4.1 The four sets of directions and exponents*

The following local sets of directions have been studied, each with *n* corresponding local exponents:

(1.1.) (A) The principal axes $\underline{a}$ in **A** of the 'full' Jacobian deformation ellipsoid **S**, with local exponents α.

(1.2.) (B) The 'constrained' flow direction $\underline{f}_\parallel$ and the extreme divergence directions $\underline{b}_\perp$ of the 'reduced' Jacobian submatrix $\mathbf{S}^{\perp(2,\ldots n)}$ (in contrast to the 'full' **S**), both described by **H**, with local exponents $\varphi_\parallel$ and $\beta_\perp$.

(2.1.) (V). The 'free' Oseledec (O) unique local frame **V** approximated after a transient time $\tau_{transV}$, with local exponents $\lambda_V$.

(2.2.) (W) The 'constrained' Rateitschak and Klages (RK) unique local frame **W** approximated after a transient time $\tau_{transW}$, with local exponents $\lambda_{\parallel W} = \varphi_\parallel$ and $\lambda_{\perp W}$.

The connections between these four local schemes are illustrated in Table 1.

| | free<br>'full' | constrained<br>$\parallel$ and<br>'reduced'<br>$\perp$ to d$\underline{x} \parallel \underline{f}_\parallel$ | |
|---|---|---|---|
| strictly local for all points $\underline{x}$ in phase space<br><br>no integration | 1.1.<br>**A**  $\underline{a}_i$<br><br>$\alpha_i$ | 1.2<br>**B**  $\underline{b}_\parallel = \underline{f}_\parallel$,<br>$\underline{b}_{\perp k}$<br><br>$\beta_\parallel = \varphi_\parallel$ and $\beta_{\perp k}$ | local extreme<br>matrix, unit vectors<br><br>local exponents |
| local at $\underline{x}(t)$, but after integration along trajectory $\{\underline{x}(t_k)\}$ $k=0\ldots n$ | 2.1<br>**V**  $\underline{v}_i$<br><br>$\lambda_{Vi}$ | 2.2<br>**W**  $\underline{w}_\parallel = \underline{f}_\parallel$,<br>$\underline{w}_{\perp k}$<br><br>$\lambda_{\parallel W} = \varphi_\parallel$ and $\lambda_{\perp Wk}$ | local 'unique'<br>matrix, unit vectors<br><br>local exponents |

**Table 1.** Illustration of the relations between the elements in the four local frames. Note that the parallel exponents of 1.2. and 2.2. are equal and both need no integration.

*4.2 Comparison of 'full' exponents and directions with the 'constrained' case*

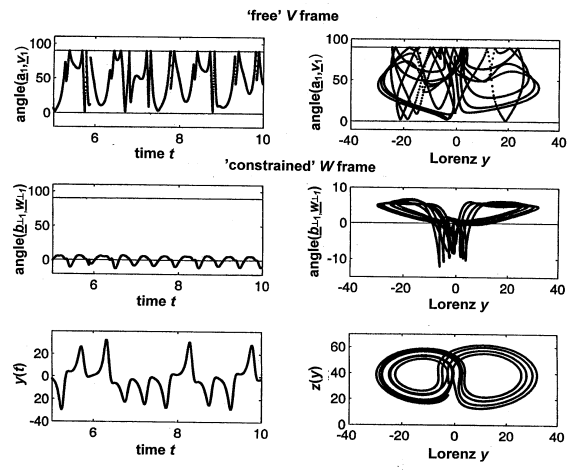

**Fig. 5.** Lorenz chaos; Bottom, *Left*: *y* component vs. time *t*. *Right*: *z* component vs. *y* component showing the two loops. Top: The angle between the principal axis $\underline{a}_l$ of the local Jacobian deformation ellipsoid **S** with the largest exponent and the first axis $\underline{v}_l$ of the **V** frame. *Left*: vs. time *t*. *Right*: vs. *y* component. Centre: The angle between the axis $\underline{b}_{\perp l}$ of the local extreme exponent orthogonal to the flow (main axis of the reduced ellipsoid $\mathbf{S}^{\perp(2,\ldots n)}$) and the first axis $\underline{w}_{\perp l}$ orthogonal to the flow of the **W** frame. *Left*: vs. time *t*. *Right*: vs. *y* component.
Note that the free **V** frame (top) has a much more complex relation to the trajectory than the constraint **W** frame (centre).

Obviously, the extreme local exponents of 'full' (α) and 'reduced' (β) frames are the limiting values of the corresponding exponents of 'free full' ($\lambda_V$) and 'constrained' ($\lambda_W$) frames, respectively. Deviations between these exponents will be large when the corresponding directions of the different frames are very different. Therefore, angles between the main directions in the different frames for the Lorenz model with $\{\underline{x}\}=\{x,y,z\}$ are shown in Fig. 5 as functions of time $t$ (left) and of the variable $y$ (right), with the variable $y(t)$ and $z(y)$ of the trajectory at the bottom. On top, the 'full' case indicates that the angle between the principal axes ($\underline{a}_1$) of the 'full' frame $A$ and the main axis ($\underline{v}_1$) of the 'free' frame $V$ is a complicated function of the trajectory.

At the centre of Fig. 5, however, the 'constrained' case is shown with the angle between the extreme orthogonal divergence direction and the main orthogonal axis of the frame $W$. This angle has only a small variation (centre left). It is also somewhat closer (centre right) to the shape of the attractor (bottom right).

### 4.3 Eliminating the ambiguity of the starting frame

In both the O and RK method the starting frames $V_0$ and $W_0$ are ambiguous. This ambiguity can be removed if the corresponding local frame defined by the Jacobian deformation ellipsoid is used. For the Oseledec frame the set $A_{\underline{x}}$ refers to the principal deformation directions. For RK the frame $H_{\underline{x}}$ has as a first vector the flow direction, the remaining ones are the local extreme directions of orthogonal divergence. For RK this setting shortens the transient time until the $W_{\underline{x}}$ frame is close to the unique local frame, see Fig. 5 centre, whereas $A_{\underline{x}}$ and $V_{\underline{x}}$ might be very different, see Fig. 5 top, resulting in a long transient.

### 4.4 Comparing the local exponents as functions of time t

The Lorenz model will be used to study local exponents as functions of time $t$.

Fig. 6 (bottom) shows the $x$ component as a function of time for a short time interval. Above, local exponents are displayed for the same time interval, the left side for the 'free' 'full' case, the right side for the 'constrained' 'reduced' case (see Table 1).

For the 'full' case, Fig. 6 (left, top) plots the first 'free' exponent $\lambda_{V1}$ together with the main principal exponent $\alpha_1$ (fine line) as the local extreme value. The first exponent $\lambda_{V1}$ is not at all times the largest and rarely has the extreme possible value of $\alpha_1$. The difference ($\lambda_{V1}- \alpha_1$), (left side, second row) has no relation to the $x$ component (bottom).

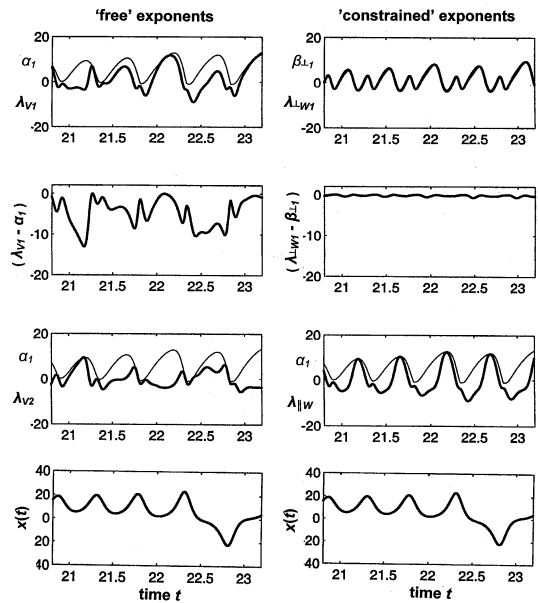

**Fig. 6.** Lorenz chaos: Local exponents vs. time $t$. <u>Bottom</u>: $x$-component. <u>Top</u> *left*: first exponent $\lambda_{V1}$ (thick line); main extreme exponent $\alpha_1$ (thin line). *Right.* first exponent $\lambda_{W1}$ and extreme local separation $\beta_{\perp 1}$ orthogonal to flow (nearly same values). <u>Second row</u>: *Left*: Difference ($\lambda_{V1}- \alpha_1$); *right:* difference ($\lambda_{\perp W1}- \beta_{\perp 1}$) at the same scale as left side. <u>Third row</u>: thin line: main extreme exponent $\alpha_1$ *Left*: second exponent $\lambda_{V2}$; *right*: second exponent $\lambda_{W2}= \lambda_{\|W}= \varphi_{\|}$ acceleration along the flow. Note: Both local exponents $\lambda_{V1}$ and $\lambda_{V2}$ (left) consist of a mixing of separation $\lambda_{\perp W1}$ and acceleration $\lambda_{\|W}$ (right).

The third row, left, displays the second 'free' exponent $\lambda_{V2}$, again with $\alpha_1$ (fine line). Remember that for $t\to\infty$ the average of the first exponent is positive and the second average approaches zero. Both local exponents $\lambda_{V1}$ and $\lambda_{V2}$ are a mixing of divergence and acceleration.

For the 'constrained' case, Fig. 6 (right), which discriminates between divergence and acceleration, Fig 6 (right, top) shows the main principal exponent orthogonal to the flow $\beta_{\perp 1,}$ thus indicating the possible maximum of divergence. The first exponent for divergence $\lambda_{\perp W1}$ is very close to that maximum $\beta_{\perp k}$, with the small difference ($\lambda_{\perp W1}- \beta_\perp$) displayed below, related to the small deviation of the relative directions, as shown in Fig. 5 centre.

However, the exponent for the local acceleration $\lambda_{\|W}= \varphi_{\|}$ of Fig. 6 right, third row, can have values nearly twice as large as the exponent $\lambda_{\perp W1}$ for divergence, but clearly never exceeds the main principal 'full' exponent $\alpha_1$ (fine line).

At this point it is interesting to ask why the global exponents are equal for both frames as demonstrated in Fig. 2, since their local behaviour is quite different. Assuming validity of the conjecture of Moser and Meier [20] that the direction for the smallest exponent is always orthogonal to d$x$ for the *V* frame and the numerical result that it is parallel to the corresponding direction of the *W* frame, both smallest exponents are equal locally and globally. Knowing further that one global exponent must be zero for the Lorenz model and that the sum of all three exponents is equal, also the largest global exponents should be equal for the *V* and *W* frames.

### 5. Local exponents as functions of phase space $R\{\underline{x}\}$ vs. $x$ component

*5.1 Why discussing local exponents as functions of phase space?*

The temporal behaviour of the four local exponents has been extensively discussed in the previous sections. For a better understanding it seems worthwhile to explore their behaviour in the phase space $R\{\underline{x}\}$.
The directions of the Oseledec *V*-frame were already shown in 3-D plots and discussed by Wolf et al. [5] and Green and Kim [7] for a short part of the Lorenz trajectory. Here, longer parts of trajectories changing the loop several times will be displayed.. First the magnitude of the exponents will be shown as a function of the x component. Then the trajectories will be shown projected onto the *yz* plane if the values of the associated different exponents exceed a certain limit *c*.

*5.2 Lorenz local exponents as a function of the x component*

Fig. 7 shows on top a time series of the *x* component of 'stable' chaos. Below, left side, the exponents of the 'free' *V* frame are displayed to be compared to the exponents of the 'constrained' *W* frame. In order to show also the variation of the speed, the figures plot points at equal time intervals. First, the largest exponent $\lambda_{V1}$ (left) has a rather complex behaviour as compared to the true divergence exponent $\lambda_{\perp Wk}$ (right). Moreover, close to the *x* value zero all values of the divergence exponent $\lambda_{\perp Wk}$ are positive, whereas the 'free' exponent $\lambda_{V1}$ has a wide spread of positive and negative values in this *x* range around zero. This range is important, since here there are only trajectories which diverge from one loop to the other loop, as is easily seen on the display on top. The reason for the spread of values of $\lambda_{V1}$ in this diverging range is its mixing of divergence with acceleration, which is strongly negative in this range. The local exponent $\lambda_{\|W} = \varphi_\|$ is shown on the second row, right, together with the second exponent $\lambda_{V2}$ left, again exhibiting a complex structure with respect to the *x* component.

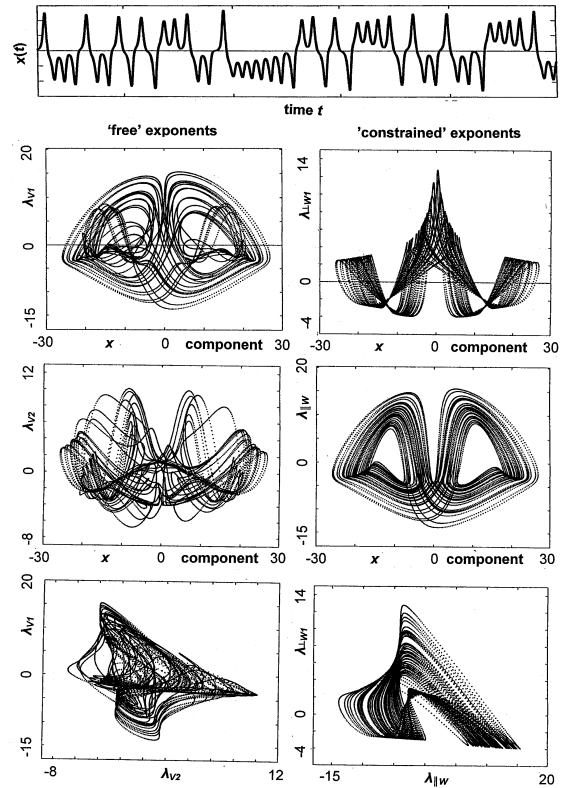

**Fig. 7.** Lorenz chaos: <u>Top</u>: *x* component vs. time *t* for a longer time interval. <u>Below</u>: Local exponents vs. *x* component for the same time interval. *Left*: *V* frame. *Right*: *W* frame. First row $\lambda_{V1}$ and $\lambda_{W1}$, below $\lambda_{V2}$ and $\lambda_{W2} = \lambda_{\|W} = \varphi_\|$. Bottom row: $\lambda_{V1}$ as a function of $\lambda_{V2}$ *(left)*, and $\lambda_{\perp Wk}$ as a function of $\lambda_{\|W} = \varphi_\|$ *(right)*.

How is the relation between the first and the second exponent? In order to indicate this relation, the row at the bottom displays $\lambda_{V1}$ as a function of $\lambda_{V2}$ (left), and $\lambda_{\perp Wk}$ as a function of $\lambda_{\|W} = \varphi_\|$ (right).
In summary, in Fig. 7 all plots of the 'free' *V* frame (left) show a rather complicated structure compared with the plots of the 'constrained' *W* frame (right). The reason for this will be illustrated in the next figure.

*5.3 Lorenz local frame angles as a function of the x component*

Figure 8 (top left) shows the angle of the first vector $\underline{v}_1$ of the 'free' *V* frame relative to the flow direction $f_\|$ as a function of the *x* component. Clearly, this angle has a wide spread between being nearly parallel and nearly antiparallel to the flow direction $f_\|$. In both these extreme cases the exponent has a strong admixing of acceleration.

In contrast, the first vector $\underline{w}\perp_1$ of the 'constrained' $W$ frame for divergence is by definition always orthogonal to the flow direction.

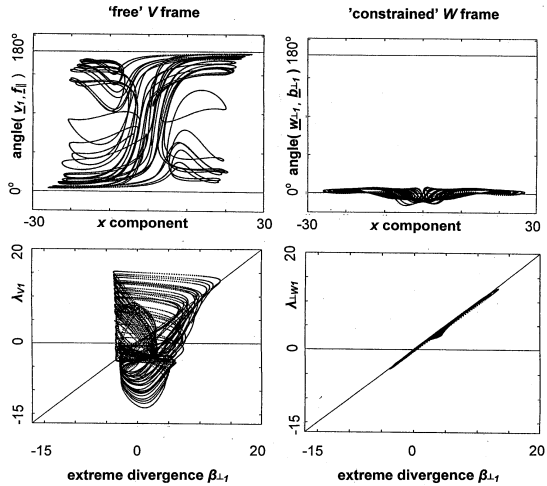

**Fig. 8.** Lorenz chaos: <u>Top left</u>: angle between first vector $\underline{v}_1$ of the 'free' $V$ frame and flow direction $\underline{f}_\|$ vs. $x$ component. <u>Top right</u>: angle between direction $\underline{w}\perp_1$ of the 'constrained' $W$ frame ($\underline{w}\perp$ always orthogonal to flow direction $\underline{f}_\|$) and direction $\underline{b}\perp_1$ of the extreme separation exponent $\beta\perp_1$ vs. $x$ component. <u>Bottom right</u>: exponent $\lambda\perp_{W1}$ vs. extreme separation exponent $\beta\perp_1$ Straight diagonal line: limit $\lambda\perp_{W1} = \beta\perp_1$. <u>Bottom left</u>: first 'free' exponent $\lambda_{V1}$ vs. $\beta\perp_1$, similar straight line $\lambda_{V1} = \beta\perp_1$.

In Fig 8 (top right) the angle of this direction $\underline{w}\perp_1$ within the orthogonal plane relative to the direction $\underline{b}\perp_1$ of the extreme divergence exponent $\beta\perp_1$ is shown to be always small as a function of the $x$ component with a range between -16 to 8 degrees. Therefore, the values of the divergence exponent $\lambda\perp_{W1}$ are never far from the value of $\beta\perp_1$.

Fig 8 (bottom right) shows the values of $\lambda\perp_{W1}$ as a function of $\beta\perp_1$ with the straight diagonal indicating the limit $\lambda\perp_{W1} = \beta\perp_1$. A similar diagonal $\lambda\perp_{W1} = \beta\perp_1$ is plotted in Fig. 8 (bottom left), showing the first 'free' exponent $\lambda_{V1}$ also as a function of $\beta\perp_1$. All the excess on the left side as compared to the right side has to be averaged out during integration in time $t$ to finally describe only divergence.

*5.4 Lorenz local exponents of integrated frames vs. local maximum exponent $\alpha_1$*

It seems worthwhile to compare in Fig. 9 the local exponents of the integrated frames $V$ (left) and $W$ (right) as functions of the strictly local maximum exponent $\alpha_1$. Obviously, $\alpha_1$ is the upper limit, which is attained at specific points of the trajectory for the first (top) and the second (bottom) exponents, indicating that the local axes of the frames coincide with the main principal axis of the Jacobi deformation ellipsoid $S$ at these specific points of the trajectory. Again the left side ($V$) has a more complex structure than the right side ($W$). An interesting question is: Where in the phase space occur the points where the exponents are close to the limiting value? A $yz$ plot reveals that about the same number of such points is widely distributed for the $V$ frames but is concentrated in a few small regions for the $W$ frame.

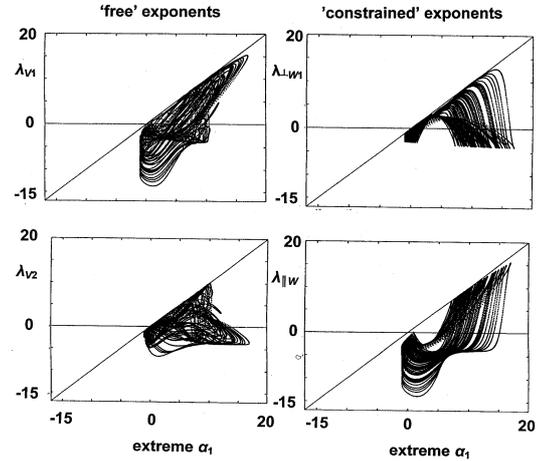

**Fig. 9.** Lorenz chaos: Local exponents of the integrated frames $V$ (left) and $W$ (right) vs. strictly local maximum exponent $\alpha_1$. Top: first exponents, bottom second exponents. Diagonal lines: exponent = $\alpha_1$ as a limit.

## 6. Local exponents and angles as functions of phase space $R\{\underline{x}\}$ in the $yz$ plane

*6.1 The local exponents as functions of phase space $R\{\underline{x}\}$ in the $yz$ plane*

Fig. 10 and Fig. 11 plot the projection of the trajectory onto the $yz$-plane when the local exponents exceed $c$ with $c = 0$ in Fig. 10 and $c = 4$ in Fig. 11 Again the trajectory is plotted at equal time intervals in order to show the change of the velocity. The small circle denotes the maximal value. On both figures, the left side displays the 'free' 'full' case, the right side the 'constrained' 'reduced' case (see table 1). The top rows show the strictly local exponents $\alpha_1$ (general maximum) and $\beta\perp_1$ (maximum of divergence only). The centre display the largest exponents of the integrated frames $V$ and $W$ as $\lambda_{V1}$ (mixing divergence and acceleration) and $\lambda\perp_{W1}$ (divergence only), respectively. The bottom rows show the second exponents $\lambda_{V2}$ (mixing acceleration and divergence) and $\lambda_{\|W} = \varphi_\|$ (acceleration only). Comparing the 'free' exponents of the left side with each other, it is obvious that only $\alpha_1$ has clear cut borders of the limit $c$, whereas centre and bottom exhibit a peculiar pattern of the limit. Furthermore, centre and bottom have regions where both exponents are above the limit.

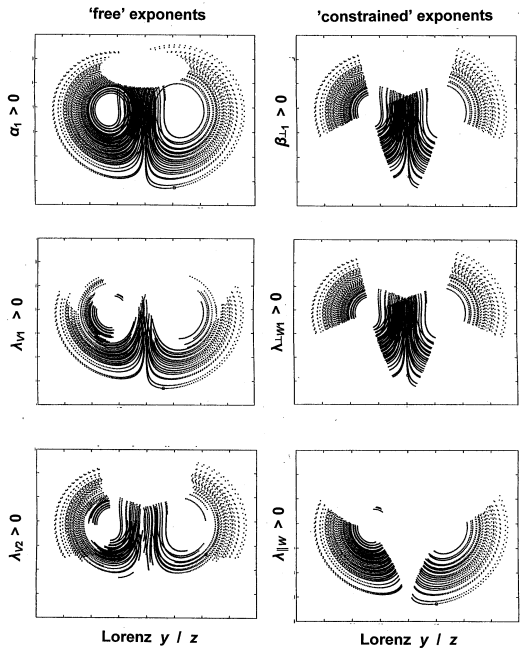

**Fig. 10**. Lorenz chaos: Projection of the trajectory at equal time intervals onto the *yz*-plane when the local exponents exceed $c = 0$. <u>Left</u>: 'free' case, <u>right</u>: 'constrained case. <u>Top</u>: maximum exponent $\alpha_1$ and maximum separation exponent $\beta_{\perp 1}$. <u>Centre</u>: first exponent $\lambda_{V1}$ and $\lambda_{\perp W1}$; <u>bottom</u>: second exponent $\lambda_{V2}$ and $\lambda_{\parallel W} = \varphi_\parallel$, respectively.

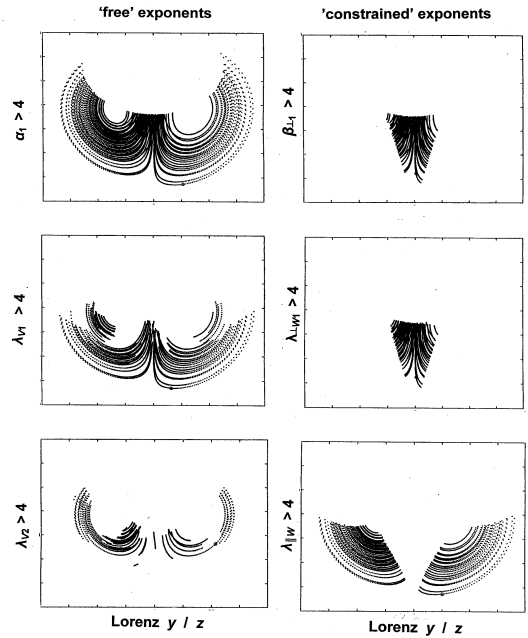

**Fig. 11.** Lorenz chaos: Similar to Fig. 10, but exceeding the limit $c = 4$.

The right side of 'constrained' exponents shows a very different behaviour. The limits are clear functions of the phase space. Moreover, the top and centre are almost similar, the centre pattern only slightly smaller. The centre and bottom patterns do not coincide.

Finally, and most importantly, only the 'constrained' case right, top and centre, has large values where the real divergence of the two loops occurs. The discrepancy to the left side is easy to understand: Since acceleration is larger than divergence, and the first exponent $\lambda_{V1}$ is a strong mixing of divergence and acceleration, the pattern bottom right $\lambda_{\parallel W} = \varphi_\parallel$ is easy to see in the pattern centre left $\lambda_{V1}$, and the maximum (small circle) is nearly at the same place. Hence, only the 'constrained' case seems to analyse directly the local chaotic behaviour.

The directions of both frames ***V*** and ***W*** are dependent on integration and not only on the location <u>x</u>. This brings about a dependence on the former sections of the trajectory. Although the exponents are unique for each location <u>x</u>, they are not simple functions of the phase space, since the former section of each location is different. The integration is a nonlinear procedure and, therefore, might have in itself a chaotic behaviour sensitive to small changes in the previous conditions. This implies a chaotic behaviour in addition to the chaotic behaviour of the analysed chaotic trajectory, resulting in the fact that another trajectory nearby might have a very different local exponent caused by a tiny difference at earlier points of that trajectory. However, this additional complexity is very different for the two frames, dependent on their 'transient times' as a measure to 'forget' earlier sections of the trajectory and on the range of changing angles of the frames. This complex behaviour is also present when angles of selected directions of these frames are plotted, as is performed in the next section.

*6.2 Local angles as functions of phase space R{<u>x</u>} in the yz plane*

Figure 12 shows angles within certain bounds in the *yz* plane: left the angle of the first vector of the 'free' ***V*** frame relative to the flow direction; right the angle between the first orthogonal vector of the 'constrained' ***W*** frame relative to the direction of the extreme divergence direction orthogonal to the flow. The same angles were shown in Fig.8, top, but only as functions of the *x* component.

The bounds imposed on the left are smaller than 10, between 60 and 120, and larger than 120 degrees, from top to bottom, respectively.

The bounds on the right are much narrower, since this angle is between -16 and -4, between -2 and 2, and between 4 and 8 degrees, from top to bottom, respectively, with the total range being from -16 to 8 degrees, see Fig. 8, top right. Since this range is small and the angle fluctuates on both sides (defined via the zero degree) of the extreme direction, the resulting exponents fluctuate much less than the exponents of the *V* frame. Furthermore, although the *V* frame changes sometimes through 90 degrees, see Fig. 12 left centre and Fig. 8 left top, and thus through the plane orthogonal to the flow, its direction within this plane has not to coincide with the orthogonal direction of the *W* frame at this point. However, since the third directions of both frames have been found to be nearly parallel, this plane is not far from the second direction of the *W* frame.

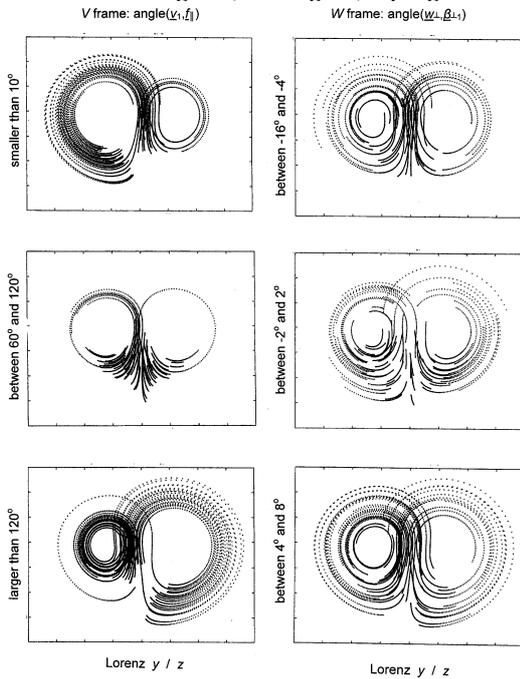

**Fig. 12.** Lorenz chaos: Projection of the trajectory at equal time intervals onto the *yz*-plane when angles are within limits. <u>Left</u>: angle between first vector $v_1$ of 'free' *V* frame and flow direction $f_∥$. The limits are smaller than 10, between 60 and 120, and larger than 120 degrees, from top to bottom, respectively. <u>Right</u>: angle between first orthogonal vector $w_{⊥1}$ of 'constrained' *W* frame and direction $b_{⊥1}$ of the extreme separation orthogonal to the flow. The limits are much narrower, since this angle is within the range from -16 to 8 degrees in the Fig. 8 top right: here between -16 and -4, between -2 and 2, and between 4 and 8 degrees from top to bottom, respectively.

## 7. Local divergence in the phase space $R\{\underline{x}\}$ outside trajectories

Since the 'constrained' exponent $\beta_{⊥1}$ of true divergence is only a function of the position $\underline{x}$ in the phase space and does, therefore, not depend on a trajectory, its value can be evaluated directly in the full phase space $R\{\underline{x}\}$. These values are shown in Fig. 13 as a mesh plot for the *xy* plane at *z* values 60 (top left) and 20 (bottom left). The strange Lorenz attractor is about along the diagonal $x ≈ y$ where the values $\beta_{⊥1}$ are low, but with a higher pass between the sections of the two loops.

A simple local indicator of the curvature of trajectories in the phase space can be found easily: For the evaluation of the local acceleration, the new vector $J f_∥$ after the time interval d$t$ is projected by the scalar product ($\underline{f}_∥ . J f_∥$) onto the flow direction $f_∥$. The absolute magnitude of the new vector minus the absolute magnitude of the projection is a measure for the deviation at t+d$t$ from the direction $f_∥$ at $t$ of the local trajectory and hence a measure of the curvature of the trajectory,

$$d = \big| J f_∥ \big| - \big| (\underline{f}_∥ . J f_∥) \big| \qquad (15)$$

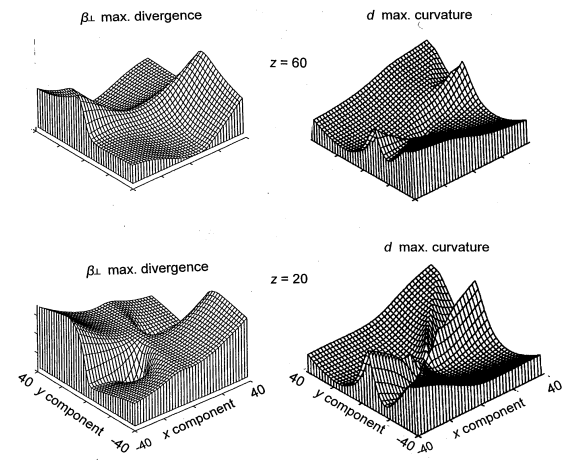

**Fig. 13.** Lorenz chaos: The 'constrained' exponent $\beta_{⊥1}$ of true separation is only a function of the position $\underline{x}$ (independent of any trajectory) in the phase space $R\{\underline{x}\}$. Its values are plotted as a mesh on an *xy* plane at *z* values 60 (top) and 20 (bottom) on the left side. The local measure *d*, eq. (15), of the curvature along the trajectory through $\underline{x}$ is plotted on the right side for the same *z* values.

This 'deviation' *d* is shown in Fig. 13 on the right side. In more detail, the Jacobian matrix *J* can be written as sum of the symmetric matrix *S* and the anti-symmetric *D*: *J* =*S*+*D*, see sect. *2.1*. The evaluated separate action of these matrices onto *d* reveals a very different behaviour: The symmetric *S* causes a strongly varying peculiar pattern in the region of the strange attractor, but its action is nearly negligible outside the attractor. In contrast, the anti-symmetric *D* creates a rather smooth structure, small at the attractor, but increasingly large with larger distance from the attractor. The combined action of these different sources are visible in Fig. 13 on the right side as varying peculiar pattern along the diagonal (**S**), and smooth increase outside (**D**).

## 8. Why integrating exponents for short intervals between Poincaré points ?

*8.1 The interest of integration during shorter time intervals than infinity*

At the beginning of analysing chaos, the main interest was to find general numbers defined in the limit of $t \to \infty$, which presupposes a 'stable' chaos. Later on, also chaotic motions were analysed where the strength of the chaotic behaviour was changing. Therefore, shorter time intervals were used. The large variation of the local exponents in time made the resulting numbers strongly dependent on the limits where the time intervals start and end. In order to test different methods, a chaotic behaviour with strongly changing character would be welcome. Such a 'transient' chaos will be described in the next section.

*8.2 'Stable' and 'transient' Lorenz chaos*

Until now, the Lorenz chaos with parameters $(\sigma,\rho,\beta) = (16,40,4)$ was 'stable' with repelling unstable fixed points at $\underline{X}=0$, where trajectories starting nearby would spiral out, join the well-known double loop strange attractor and stay there in theory to infinity, in practice until the build-up of computing errors will be too high. The top of Figure 14 displays a section of the 'stable' chaos where the trajectories are plotted on the *yz* plane at equal time intervals to show the variation of the speed as in earlier plots.

A change of the parameter $\rho$ in the Lorenz equation ($dy/dt = -xz + \rho x - y$) from 40 to 28.165 replaces the unstable fixed points by attracting stable fixed points. Figure 14 bottom displays a peculiar 'transient' Lorenz trajectory starting with several loops on the right side with increasing radius, followed by a chaotic interval with loops on both sides, and a final decay spiralling on the left side to a stable fixed point.

*8.3 Time intervals $\tau_p$ between Poincaré points for Lorenz trajectories*

For the Lorenz chaos, integrating over one loop would be reasonable. Since there are no closed loops, well chosen Poincaré points will be used to define the start and end of a single loop. Poincaré points will be defined here when a trajectory is crossing a plane at $y=c_p$ 'from above', i.e. for decreasing values of *y*. The time interval $\tau_p$ is then defined between two consecutive Poincaré points. The plane parameter $c_p$ is chosen at a level where the trajectories are about normal to the Poincaré plane. Figure 14 top shows this plane ($c_p=40$) as a line in the *yz* plot for a 'stable chaos' with instable fixed points, whereas Fig. 14 bottom displays a similar line ($c_p=27.165$) for stable attracting fixed points with the 'transient' trajectory described above.

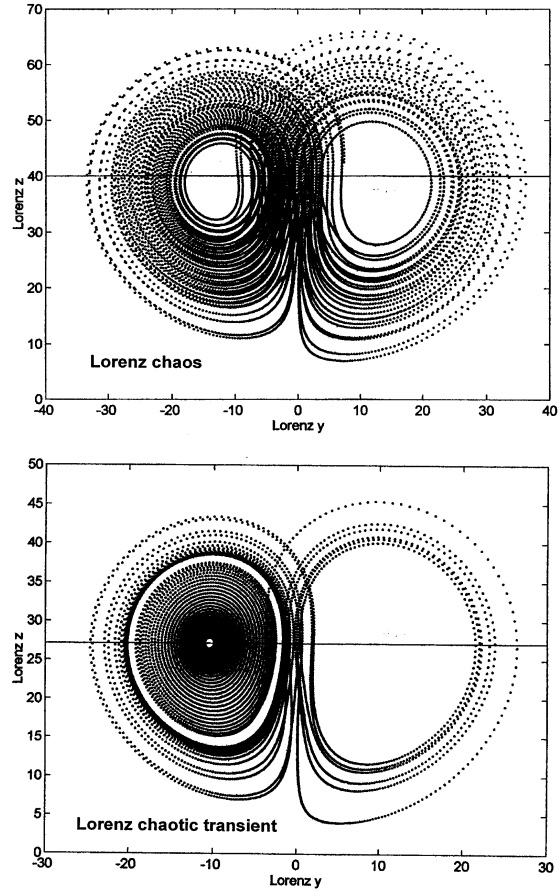

**Fig. 14.** Lorenz model: 'stable chaos' (above) and 'chaotic transient to a fixed point' (bottom). Trajectories are projected onto the *yz* plane at equal time intervals to indicate the changes in velocity. The lines indicate the position of the Poincaré planes $z=c_p$=const. with $c_p=40$ (above) and $c_p=27.165$ (below). Over each time interval between consecutive Poincaré points local exponents are integrated.

The integration between Poincaré points is therefore performed over one loop. The results are shown in the next sections for the 'free' and 'constraint' case, first for the 'stable' Lorenz chaos. These results will be compared with the integration of all possible local directions, which are evaluated with the local Jacobian deformation ellipsoid.

## 9. The idea of integrating all directions of the local sphere

*9.1 Description of the idea of integrating all directions*

It seems worthwhile to recall here the basic idea behind the definition of Lyapunov exponents, which consists of integrating along a trajectory the subsequent deformations of the surface of a sphere, surrounding a given initial condition, along all direction in phase space. At the final time, the principal axes and eigenvalues of this deformed ellipsoid are considered to be the relevant characteristics of the problem, thus only $n$ exponents are sufficient to describe the type of dynamical instability. It was proved by Oseledec that only the knowledge of the deformations of a frame of $n$ orthogonal directions is necessary to find these final axes. The evaluation of the deformation with the Jacobian ellipsoid $S$ enables to test this established fact numerically.

*9.2 The problem of subsequent rotation*

It is straightforward to evaluate $S[\underline{x}(t)]$ along a trajectory $\underline{x}(t)$. However, for calculating Lyapunov exponents from it the rotations of the directions after each time step $dt$ still have to be eliminated. This important fact is not too obvious, since the expression $J\underline{u}$ of eq. (3) implies elongation and rotation. However, in order to find instantaneous local exponents describing only elongation, the rotation must be projected out by using the scalar product of eq. (5). Moreover, because this scalar product has the same vector on both sides, it enforces the symmetrisation of the Jacobian matrix $J$ in form of the symmetric Jacobian deformation ellipsoid $S$. Therefore, a rotation of $S$ has to be performed after each time step $dt$ along the trajectory. In principle, each direction should be rotated according to eq. (1) separately. However, as a crude oversimplification all orientations might be rotated equally according to the rotation of an $n$ dimensional orthogonal frame evaluated with the equation of motion eq. (1). Along these lines, transformations $S$ into the frames $V$ or $W$ will be used. Again, the average of values computed at discrete time steps will approximate the continuous time integration. Let the interval $\tau$ be divided into $m$ parts $\Delta t$. With $\underline{x}_k = \underline{x}(t_0 + k\,\Delta t)$, and first for the frame $V$, the resulting arithmetic average $S^{(\tau)}_V$ can be written as

$$S^{(\tau)}_V = (1/\tau)\sum_{k=1}^{m} V^T(\underline{x}_k)\ S_x(\underline{x}_k)\ V(\underline{x}_k) \qquad (16)$$

Secondly, in complete analogy the average $S^{(\tau)}_W$ is defined for the RK frame $W$ with the first vector constrained along the flow direction,

$$S^{(\tau)}_W = (1/\tau)\sum_{k=1}^{m} W^T(\underline{x}_k)\ S_x(\underline{x}_k)\ W(\underline{x}_k) \qquad (17)$$

Thirdly, another approach consists of considering only all the directions orthogonal to the flow as described by the reduced Jacobian ellipsoid $S\perp^{(2,\ldots n)}$, obviously to be transformed by a reduced matrix $W\perp^{(2,\ldots n)}$, resulting in the average $S\perp^{(\tau)}$,

$$S\perp_W^{(\tau)} = (1/\tau)\sum_{k=1}^{m} W\perp^T(\underline{x}_k)\ S_x(\underline{x}_k)\ W\perp(\underline{x}_k) \qquad (18)$$

At the final time of the integration, for each of the above averages the eigenvalues are evaluated and compared with the integrals of the exponents of the corresponding frames. For convenience, the eigenvalues will be ordered with the largest first for the principal axes of the averaged sums of the ellipsoids.

In contrast, the averaged sums of the local exponents of the frames are ordered according to the order during the orthogonal normalization process, which might result in a reversed order, such that the value for the first exponent is smaller than for the second exponent.

## 10. Integration of all directions and integration of $n$ local exponents for 'stable' chaos

*10.1 Integrations between Poincaré points*

Figure 15 shows for 'stable' Lorenz chaos, see the variable $x(t)$ at the bottom, the resulting first and second exponents, evaluated as eigenvalues of the final Poincaré integration matrix $S^{(\tau)}_V$ (top) of eq. (16), $S^{(\tau)}_W$ (second row) of eq. (17), and $S\perp_W^{(\tau)}$ (third row) of eq. (18), see symbols o and *, respectively. They are compared to the corresponding integrals of the first two exponents, evaluated as averages of the instantaneous local exponents in the directions of the frames $V$ and $W$, see symbols x and + in the figure, respectively.

For the first case $S^{(\tau)}_V$ (top), large discrepancies are visible. For the second case $S^{(\tau)}_W$ (second row), there exist rather small deviations. Only for the third case $S\perp_W^{(\tau)}$ (third row), the eigenvalues of the integral of the reduced deformation ellipsoid give the same values as the integral of the instantaneous local exponents of the $W\perp$ frame. The discrepancies might be caused by the oversimplification of rotating all orientations equally. However, the most important result is that the RK method yields values that are equivalent to the method based on computing eigenvalues by integrating all directions in the *n*-1 space orthogonal to the flow, even when these orientations are rotated equally but within the orthogonal subspace. The second exponent is by definition equal for RK and $S_{\|W}^{(\tau)}$. The small but distinct discrepancies between the full $S^{(\tau)}{}_W$ (second row) and $S_{\perp W}^{(\tau)}$ (third row) are easy to understand: Already at each point $\underline{x}$ the eigenvectors of the reduced matrix $S_\perp$ and the direction of the flow $d\underline{x}$ only accidentally coincide with the principal axes of the full local $S$

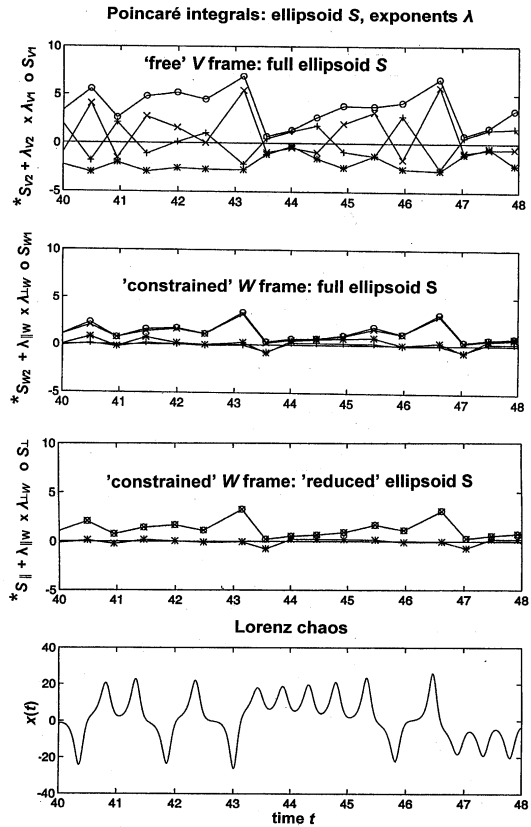

**Fig. 15.** 'Stable' Lorenz chaos; Bottom: $x(t)$. Above: Integrals of deformations between consecutive Poincaré points and subsequent eigenvalues for $S^{(\tau)}{}_V$ (top), $S^{(\tau)}{}_W$ (second row) and $S_{\perp W}^{(\tau)}$ (third row) as o and *, respectively, together with the integrals of the first two exponents of the frames $V$ and $W$ as x and +, respectively. Lines are only guides to the eye.

*10.2 Integrations of all deformations for long times compared to global exponents*

The same comparison is made after a long integration time $\tau$ when the sums are close to the asymptotic values. First, using eq. (16), the ellipsoid $S^{(\tau)}{}_V$ of the Oseledec type $V$ frame of free rotations has the diagonal elements (1.59, 0.002, -22.59). According to the conjecture for Lyapunov exponents, the global ellipsoid should have principal axes parallel to the directions of the $V$ frame. Equally important, the eigenvalues of $S^{(\tau)}{}_V$ should be equal to the global exponents evaluated by the integration of the instantaneous local exponents in the local directions of the frame $V$, resulting in the values (1.38, 0.002, -22.38). Indeed, there is a small but distinct difference for two values. Moreover, the directions of the principal axes of $S^{(\tau)}{}_V$ deviate by about 5 degrees from the directions of the $V$ frame.

Secondly, using eq. (17) the ellipsoid $S^{(\tau)}{}_W$ has the eigenvalues (1.37, 0.0001, -22.37), exactly as the corresponding global exponents of the $W$ frame (1.37, 0.0001, -22.37). Moreover, the corresponding directions differ by less than 1 degree. Note that for short intervals as described in the previous section, for $S^{(\tau)}{}_W$ the corresponding values are slightly different. Hence, for the global values these small discrepancies are levelled out for long integration time $\tau$.

At this point it seems interesting to discuss the question: Are the global 'Lyapunov eigenvectors' fixed in time? The answer is *yes* and *no*, depending on the observer. For a static observer at the origin of the coordinate system of $R\{\underline{x}\}$, the eigenvectors are fixed only if no rotating system $U(t)$ is used. In reality, these eigenvectors are rotating as fast as $U(t)$ itself. Only a rotating observer at the origin of $U(t)$ sees the Lyapunov eigenvectors fixed in time. Moreover, although the final ellipsoid is fixed accordingly for this rotating observer, he would recognise that even after a very large interval $\tau$ the instantaneous local ellipsoid is still deforming as rapidly as after a short time.

In conclusion, the approach by Lyapunov as already been described in sect. 1.2 (iii) proposes that the global exponents evaluated only for *n* local orthogonal directions are the eigenvalues of the global deformation ellipsoid, which contains all directions. This idea has been tested numerically. However, it is only confirmed if the constrained frame $W$ of RK is used.

## 11. Poincaré integrals for 'transient' chaos with a sudden flip of the frames

*11.1 'Transient' with a sudden change of the frame: Oseledec V frame*

Figure 16 bottom shows the $x$ component vs. time $t$ for the 'transient' chaos described above. On top, the Poincaré integrals for the first two exponents of the $V$ frame are plotted as o and +, respectively. The centre shows the angle of the first axis of the $V$ frame relative to the flow direction. The frame starts with an arbitrary orientation. Then, at $t = 7$ the frame is suddenly rotated by 90 degrees. After this flip the reorientation to being nearly parallel to the flow is rather slow. During the short strongly chaotic regime the angle flips up and down. Long after the beginning of the spiralling down to the fixed point the angle reorients again between $t \approx 22$ and 27. This delayed motion causes a virtually chaotic pattern of the exponents, see top panel within approximately the same time interval, similar to the reaction at the beginning and after the flip. Furthermore, note that during the strong chaotic behaviour the integral of the second exponent is sometimes larger than the first exponent.

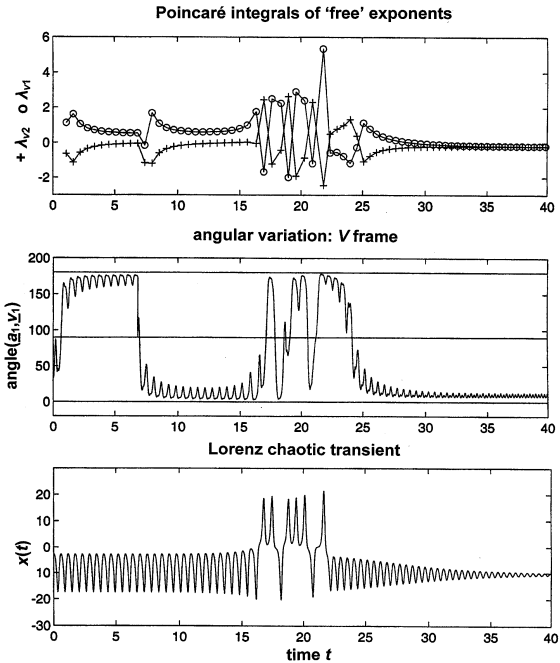

**Fig. 16.** Lorenz 'chaotic transient to a fixed point'; Bottom: $x(t)$. Centre: Angle of the first axis of the *V* frame relative to the flow direction. This angle is artificially rotated by 90 degrees at $t = 7$. Top: Poincaré integrals of the first two exponents of the *V* frame denoted by the symbols o and +, respectively. Lines are only guides to the eye. Note the 'artificial bump' between $t \approx 22$ and 27 as explained in the text.

Clearly, the Oseledec *V* frame has problems with delayed reorientation, which could give wrong exponents for finite time intervals.

### 11.2 'Transient' with a sudden change of the frame: RK *W* frame

Figure 17 bottom displays the $x$ component of the same 'transient' chaos as in Fig. 16. Here the frame is also suddenly rotated by 90 degrees at $t=7$, as shown in the centre for the angle between the first orthogonal direction of the *W* frame and the direction of the local orthogonal direction of extreme divergence. This flip changes the exponents only for the integral over one loop, as is shown on top for the second exponent, since the recovery time of the angle is small, and one direction along the flow remains fixed according to the definition of the 'constrained' *W* frame. This implies that the direction for the first exponent is independent of the orientation of the other directions of the *W* frame, so the flip at $t=7$ does not affect the first exponent, see + in Fig. 17 top panel,

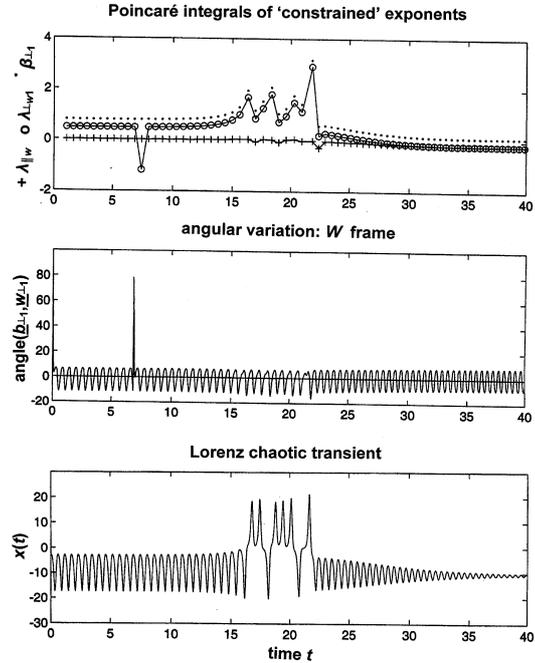

**Fig. 17**. Lorenz 'chaotic transient to a fixed point'; Bottom: $x(t)$. Centre: Angle between the first orthogonal direction of the *W* frame and the direction of the local orthogonal direction of extreme separation. The *W* frame is artificially rotated by 90 degrees at $t = 7$. Top: Poincaré integrals of the first two exponents of the *W* frame denoted by the symbols o and +, respectively. The dots plotted above the o symbols are the integrals of the local extreme orthogonal exponents, which are independent of orientation and flipping of the *W* frame.

### 11.3 Integrating the local extreme divergence exponents

It is also interesting to compare the RK Poincaré integrals with the integrals of the local extreme orthogonal exponents evaluated as largest eigenvalues of the local reduced Jacobian ellipsoid, see the dots above the symbols for the second exponent in Fig. 17. Since these integrals are for local exponents, they are independent of the orientation of any frame. Their values are always larger than the RK values, but they provide also an excellent description of the trajectory. They are small during the last part of the spiralling dynamics to a fixed point but still slightly positive there, which indicates chaotic behaviour. However, the negative integral of the exponent along the acceleration simultaneously indicates non-chaotic behaviour.

Hence the combination of both the local extreme exponent with the exponent along the acceleration leads already to a good estimate of the type of regime, chaotic or non-chaotic. Furthermore, both exponents have zero transient time, they are correct already at the staring time $t_0$, Since no frame *V* or *W* is needed, the ambiguity of a starting frame, their integration and their transient times are avoided.

## 12. Quantitative analysis using Poincaré point correlations

### 12.1 Purpose of the method

Qualitatively, the previous figures seem to favour the RK method: The structure of the time-dependent dynamics of the exponents is simpler than the one of the established Oseledec method. Hence, it seems worthwhile to quantitatively compare both methods. There are ways to test the complexity of their local exponents, see Chlouverakis et al. [20]. However, the local exponents themselves are merely tools to analyse the chaos of a given nonlinear dynamics. The question is therefore rather: How well are the given trajectories assessed by the local exponents? How well are the exponents correlated to the trajectories? How can the degree of the associated complexity be evaluated? A method similar to an established method to compare the volatility of financial values will be constructed, namely the evaluation of the standard deviation of the distribution of the values.

### 12.2 Description of the 'Poincaré point correlated deviation' (PPCD)

It seems sufficient to analyse only a typical set of points in a reduced *n-1* dimensional subspace. For Lorenz chaos, an *xy*-plane will be defined by *z*=constant. Pairs of neighbours of the points in this plane are selected and their distance *r* is evaluated. Then, the absolute difference $\Delta\lambda$ of the corresponding local exponents is evaluated as a function of the distance *r* of these neighbouring points *i,k*.

$$\Delta\lambda_{ik}(r) = |\lambda_i - \lambda_k|$$
$$\text{with } r^2 = (x_i-x_k)^2 + (y_i-y_k)^2 \quad (19)$$

Finally, plots of the strictly local exponents are compared to the exponents of the *V* and *W* frames which incorporate integrals over previous times. This comparison can be performed in a quantitative way and results in a measure of how well the *V* and *W* frames correspond to the trajectories to be analysed. The strength of chaotic behaviour is manifest in the 'pseudo'-irregularity of the differences $\Delta\lambda_{ik}(r)$, which is measurable by its standard deviation, after a possible underlying regular behaviour has been eliminated (we call it 'pseudo'-irregular, since all data points are regularly determined by eq. (1), including the rotations of the frames).

The following quantitative procedure will be used to evaluate the distribution of the data depending on the function $\Delta\lambda_{ik}(r<r_m)$ up to radius $r_m$. Since there might be a systematic bias as a function of *r*, it seems appropriate to first get rid of this bias. Therefore, the values $\Delta\lambda_{ik}(r)$ are fitted by a linear approximation resulting in $a_{ik}(r)$ for each point *ik*. Then the standard deviation *Q* of the resulting difference $[\Delta\lambda_{ik}(r) - a_{ik}(r)]$ is considered to be an adequate measure for the correlation of the exponent $\lambda$ with the trajectories, called here 'Poincaré point correlated deviation' (PPCD),

$$Q_{\text{PPCD}}(\lambda) = \text{std}\,[\Delta\lambda_{ik}(r) - a_{ik}(r)] \quad (20)$$

The quantity $Q_{\text{PPCD}}(\lambda_{local})$ of the strictly local exponents serves as a basis to assess the strength of the chaos of the trajectory. Then, $Q_{\text{PPCD}}(\lambda_{V\text{-frame}})$ and $Q_{\text{PPCD}}(\lambda_{W\text{-frame}})$ are compared to $Q_{\text{PPCD}}(\lambda_{local})$. The resulting fractions $f_V$ and $f_V$ analyse quantitatively the correlation of both frames with the trajectories,

$$f_V = Q_{\text{PPCD}}(\lambda_{V\text{-frame}}) / Q_{\text{PPCD}}(\lambda_{local}) \quad (21)$$

$$f_W = Q_{\text{PPCD}}(\lambda_{W\text{-frame}}) / Q_{\text{PPCD}}(\lambda_{local}) \quad (22)$$
.

If $f_{\text{frame}} \gg 1$, the respective frame adds chaotic behaviour to the complexity of the trajectories, which has to be eliminated by long-time integration. For $f_{\text{frame}} \approx 1$, this frame could also be used to analyse short portions of different chaotic transients.

### 12.3 The 'Poincaré point correlated deviations' (PPCD) for Lorenz chaos

#### 12.3.1 A peculiar structure of the points in the Poincaré plane

For Lorenz chaos, fixing z=40 defines an *xy*-plane, which serves for finding Poincaré points where trajectories cross the plane. Such Poincaré points $p(t)$ have already been used for the previous 'Poincaré integrals', defined in terms of time limits between consecutive points for decreasing *z*-values. Here, both points $p^{(+)}(t)$ and $p^{(-)}(t)$ for increasing and decreasing *z*, respectively, are evaluated. However, in addition to the times $t_i$ also the positions $x_i, y_i$ are considered. A plot of these Poincaré points in the Poincaré *xy*-plane reveals a particular structure, see Fig. 18 (top): There are two branches of $p^{(-)}$ points around the centre, one with y>x, denoted by c1, and the other with *y<x*, denoted by c2. The $p^{(+)}$ points have branches on the left ($\ell$) torus and on the right (r) torus. Each branch consists of a narrow band of points with only a small scattering around a smooth curve. It is now straightforward to determine for each point $p^{(-)}(t_i)$ if the corresponding trajectory comes from $\ell$ or r and goes to $\ell$ or r by analysing the time series $p^{(+)}(t_{i-1})$, $p^{(-)}(t_i)$, $p^{(+)}(t_{i+1})$. The result is striking: the points c1 display only the sequences $\ell\to\ell$ or $\ell\to$r plotted in Fig. 18 (bottom) as circles ○ and x signs, respectively. Moreover, these two possibilities are separated in the Poincaré plane. Similarly, the c2 points consist only of an r→r or r→$\ell$ sequences, plotted as stars * and + signs, respectively. This symmetry allows restricting the analysis to the c1 branch.

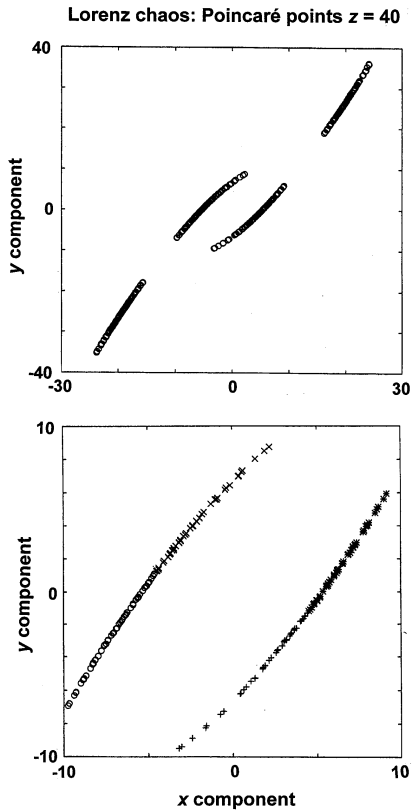

**Fig. 18.** Lorenz chaos: *xy* plot of Poincaré points at *z*=40. Top: all points. Bottom: blowup of the central sections for decreasing *z* values. Upper left branch c1: symbol o denotes the sequence $\ell\to\ell$(left torus), symbol + denotes the sequence $\ell\to$r(right torus). Lower right branch c2: * for r→r and x for r→$\ell$.

At this point an interesting question can be asked: Where on the trajectory could additional noise most easily change the dynamical behaviour? As an example, perturb the trajectory at *z*=40 where the $\ell\to\ell$ sequence is close to the $\ell\to$r sequence, thus the circles ○ and x are very close, such that the trajectory jumps from one to the other sequence. At this position a noise-induced transition as described by Gassmann [21] would be most easily possible

To provide an overview, the local exponents of the c1 points are plotted as a function of the *x* component in Fig. 19. The separation between the $\ell\to\ell$ and $\ell\to$r sequences is marked by a vertical line. The strictly local exponents are: extreme expansion $\alpha_1$ (top left), extreme orthogonal divergence $\beta_1$ (top right). The second exponent $\lambda_{\|W}$ of the *W* frame of local acceleration is also a strictly local exponent (bottom right). The central row displays the first exponents $\lambda_{V1}$ and $\lambda\perp_W$, the bottom row the second exponents $\lambda_{V2}$ and $\lambda_{\|W}$ of the *V* frame (left) and of the *W* frame (right), respectively. It is interesting that all three plots on the right side (extreme orthogonal divergence $\beta_1$ and *W* frame) share the same feature: The values of the exponents all differ for the two sequences, the barrier is marked by a horizontal line indicating a complete correlation to the behaviour of remaining on the same left loop, or changing the loop, which certainly implies a larger divergence. This feature is missing for both exponents of the *V* frame (left centre and left bottom) showing a large spreading.

*12.3.2 The PPCD analysis of the Lorenz chaos*

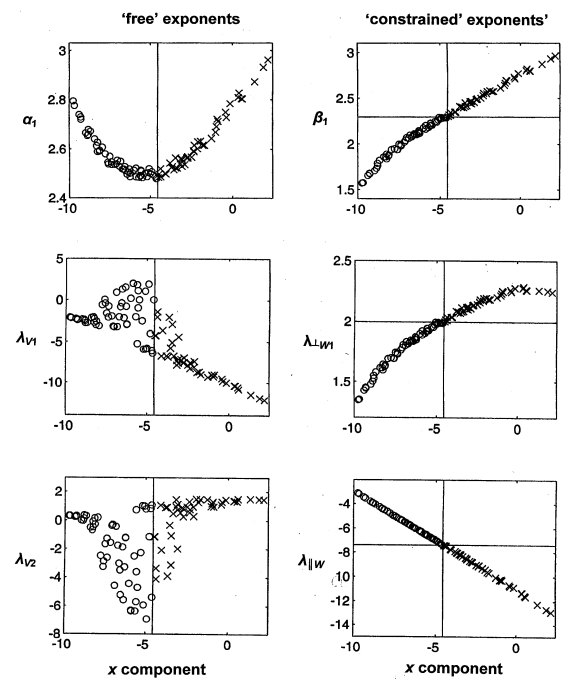

**Fig. 19.** Lorenz chaos: Local exponents of c1 points $\ell\to\ell$ (o) and $\ell\to$r (+) of Fig. 18 vs. the *x* component. Vertical lines separate o from +. <u>Left</u>: top extreme expansion $\alpha_i$, centre $\lambda_{V1}$, bottom $\lambda_{V2}$. <u>Right</u>: top extreme orthogonal expansion $\beta\perp_l$, centre $\lambda\perp_{W1}$, bottom $\lambda_{\|W}=\varphi_\|$. Horizontal lines separate values of o from +.

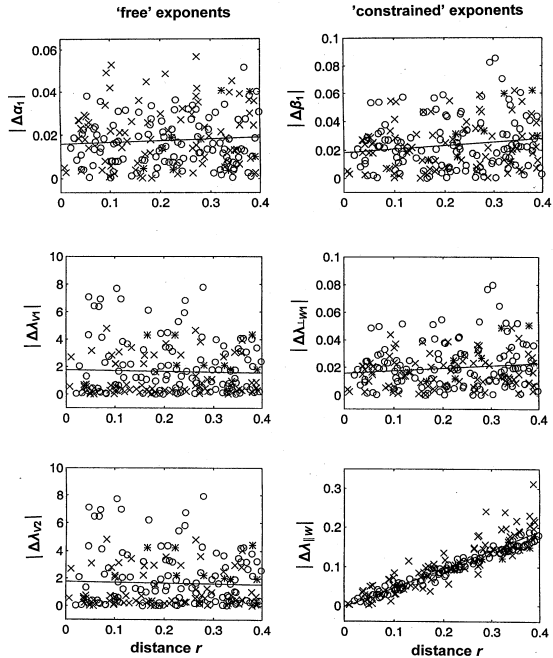

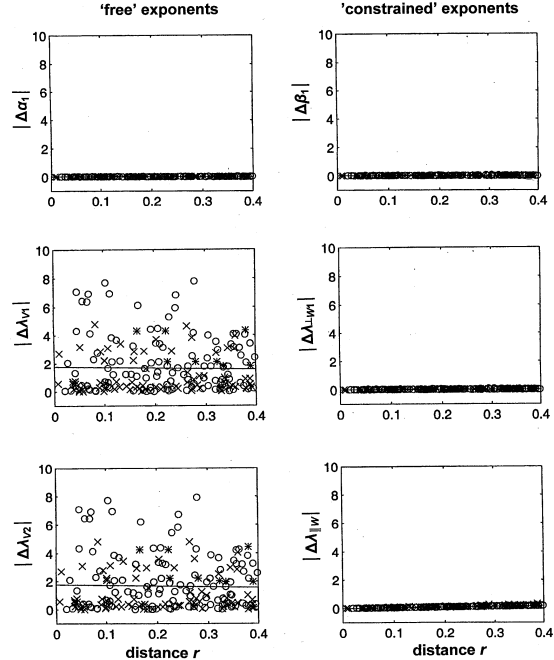

**Fig. 20.** Lorenz chaos, PPCD analysis: Absolute difference $|\Delta\lambda_{ik}(r)|$ of the corresponding local exponents $\lambda$ vs. radius $r<r_{max}=0.4$ of pairs $ik$ between $\ell \to \ell$, denoted by the symbol o, between $\ell \to r$, denoted by +, and the rare pairs between $\ell \to \ell$ and $\ell \to r$, denoted by *. Lines: linear approximations of the data points. The differences are of the same exponents as plotted in Fig. 19.

For each point of the c1 branch all local exponents are stored. Then the distance to neighbouring points is evaluated and registered for the radius $r<r_{max}=0.4$, together with the absolute difference $|\Delta\lambda_{ik}(r)|$ of the corresponding local exponents $\lambda$. These values are plotted as functions of $r$ in Fig. 20, arranged similarly to Fig. 19. The pairs $ik$ between $\ell \to \ell$ points are depicted as circles ○, pairs between $\ell \to r$ as x symbol and the rare pairs between $\ell \to \ell$ and $\ell \to r$ as stars *. The straight lines in Fig. 20 represent linear approximations to the data points. Only the differences of the acceleration $|\Delta\lambda_{\parallel W}|$ (bottom right) exhibit a distinct bias as a function of the radius $r$, due to the strong dependence of these local exponents along the branch c1, as seen for $\lambda_{\parallel W}$ in Fig. 19 (bottom right). Clearly, all plots of Fig. 20 show a strongly irregular pattern of the data. However, the magnitude of this spread is very different, see the corresponding scales, resulting in the following $Q_{PPCD}(\lambda)$ values: left, from top to bottom: 0.013, 1.8, 1.8, and right, from top to bottom: 0.017, 0.015, 0.022. The average for the three strictly local exponents (0.013, 0.017, 0.022) is 0.017. The average of the two $V$ exponents is 1.8, and the only $W$ exponent $|\Delta\lambda_{\perp W}|$ subject to integration (centre right) has a value of 0.015.

**Fig. 21.** Lorenz chaos: The same PPCD data points as shown in Fig. 20, but all plotted on the same scale to indicate the differences of the magnitude of the $V$ frame (left centre and bottom) to the $W$ frame (right centre) and to the strictly local exponents.

Therefore, the quantitative factors $f$ defined by eqs. (21) and (22) are

$$f_V \approx 100 \quad (23)$$

$$f_W \approx 1 \quad (24)$$

Although the heuristic PPCD analysis provides a strong oversimplification by using only a small selected portion of the data, its outcome is surprisingly clear.

In order to visualize this final result, Fig. 20 is potted again as Fig. 21, but now with the same scale for all plots. Obviously, only the Oseledec $V$ frame (left centre and bottom) is adding a strong chaotic complexity to the complexity of the trajectories to be analyzed.

## 13. Conclusions

This work suggests that the evaluation of Lyapunov exponents should be revisited. The introduction of the symmetric deformation Jacobian ellipsoid and its submatrix orthogonal to the flow allows the direct determination of the principal exponents and of the extreme local exponent for diverging trajectories for every point in the phase space without the need to integrate along a specific trajectory to find the local

'Lyapunov directions' according to the procedure of Oseledec. Moreover, a fundamentally different approach evaluates the divergence of nearby trajectories by avoiding a mixing with local acceleration. This is performed by adding a simple constraint to the Oseledec frame as already done in Ref. [10]: The first vector remains fixed along the flow direction d$x$, and any further vectors are subsequently orthogonalised after each rotation, which implies that only the divergence and not any partial acceleration are assessed at every point on the trajectory. This avoids the additional complexity introduced by the Oseledec method, and the local exponents are in accord with the local divergence of the trajectories. The main problem of dynamical instability: *Remain neighbouring trajectories in a tube?* is thus solved. The largest exponent indicates locally already where trajectories diverge and where the origins of instability occur,

Only the evaluation of the Jacobian deformation ellipsoid is treated analytically, the remaining conclusions are based on studying numerical examples. Hence, a formal revision of the original idea of Lyapunov to evaluate the divergence of nearby trajectories instead of considering neighbouring points remains to be carried out.

We finally remark that there might exist interesting crosslinks between our work and the very recent active field of calculating and understanding so-called Lyapunov modes in interacting many-particle systems, see Ref. [22] for a short review and further references therein. Lyapunov modes refer to the eigenmodes associated with the spectrum of Lyapunov exponents which are closest to zero, projected onto the single particles from which they originate. They were found to form interesting spatio-temporal periodic patterns. The methods developed in our paper, based on solving the eigenvalue problem for the local Jacobian deformation ellipsoid in a suitable local coordinate system, could possibly serve for developing alternative techniques of computing such Lyapunov modes. Using our methods, it might also be interesting to check for spatio-temporal structures in the corresponding distribution of local Lyapunov exponents in such systems.

**Acknowledgements**

F.W. would like to thank P.F. Meier, H.R. Moser, and E.P. Stoll for valuable help.